\documentclass[10pt,pre,twocolumn,tightenlines,longbibliography,notitlepage,superscriptaddress,nofootinbib]{revtex4-2}
\pdfoutput=1

\usepackage{xcolor}
\usepackage{verbatim}
\usepackage{graphicx}
\usepackage{natbib}
\usepackage{siunitx}
\usepackage{float}
\usepackage{xr}

\usepackage[textsize=tiny]{todonotes}

\makeatletter
\newcommand*{\addFileDependency}[1]{
  \typeout{(#1)}
  \@addtofilelist{#1}
  \IfFileExists{#1}{}{\typeout{No file #1.}}
}
\makeatother

\newcommand*{\myexternaldocument}[1]{%
    \externaldocument{#1}%
    \addFileDependency{#1.tex}%
    \addFileDependency{#1.aux}%
}
\myexternaldocument{supplemental}

\newcommand*{\rom}[1]{\uppercase\expandafter{\romannumeral #1\relax}}

\begin{document}

\title{Symmetry-guided inverse design of self-assembling multiscale DNA origami tilings}

\author{Daichi Hayakawa}
\affiliation{Martin A. Fisher School of Physics, Brandeis University, Waltham, Massachusetts 02453, USA}
\author{Thomas E. Videb\ae k}
\affiliation{Martin A. Fisher School of Physics, Brandeis University, Waltham, Massachusetts 02453, USA}
\author{Gregory M. Grason}
\affiliation{Department of Polymer Science and Engineering, University of Massachusetts, Amherst, Massachusetts 01003, USA}
\author{W. Benjamin Rogers}
\email{wrogers@brandeis.edu}
\affiliation{Martin A. Fisher School of Physics, Brandeis University, Waltham, Massachusetts 02453, USA}

\begin{abstract}
Recent advances enable the creation of nanoscale building blocks with complex geometries and interaction specificities for self-assembly. This nearly boundless design space necessitates design principles for defining the mutual interactions between multiple particle species to target a user-specified complex structure or pattern.
In this article, we develop a symmetry-based method to generate the interaction matrices that specify the assembly of two-dimensional tilings which we illustrate using equilateral triangles.
By exploiting the allowed 2D symmetries, we develop an algorithmic approach by which any periodic 2D tiling can be generated from an arbitrarily large number of subunit species, notably addressing an unmet challenge of engineering 2D crystals with periodicities that can be arbitrarily larger than subunit size.
To demonstrate the utility of our design approach, we encode specific interactions between triangular subunits synthesized by DNA origami and show that we can guide their self-assembly into tilings with a wide variety of symmetries, using up to 12 unique species of triangles. 
By conjugating specific triangles with gold nanoparticles, we fabricate gold-nanoparticle supercrystals whose lattice parameter spans up to 300 nm.
Finally, to generate economical design rules, we compare the design economy of various tilings. 
In particular, we show that (1) higher symmetries allow assembly of larger unit cells with fewer subunits and (2) linear supercrystals can be designed more economically using linear primitive unit cells.
This work provides a simple algorithmic approach to designing periodic assemblies, which may open new doors to the multiscale assembly of superlattices of nanostructured ``metatoms" with engineered plasmonic functions.

\end{abstract}

\maketitle
Self-assembly is a powerful method for building ordered structures using components ranging in size from nanometers to micrometers \cite{Whitesides2002Mar}. Unlike conventional `top-down' manufacturing, in self-assembly, the instructions for building a final material structure are encoded in the geometry and the interaction specificity of the individual building blocks. In the past few decades, various techniques for synthesizing nanoscale building blocks have been developed, such as DNA-grafted colloids \cite{Mirkin1996Aug, Macfarlane2011Oct, wang2015crystallization, he_colloidal_2020}, DNA origami \cite{rothemund_folding_2006, gerling_dynamic_2015, Tikhomirov2017MonaLisa, Wang2021May}, DNA tiles and bricks \cite{wei_complex_2012, ke_three-dimensional_2012, ke2014dna, ong2017programmable}, and de-novo protein design \cite{bale_accurate_2016, shen_novo_2018, Sahtoe2022Jan}. This ever-expanding suite of user-prescribed building blocks has enabled the self-assembly of increasingly complex architectures and devices, including crystals \cite{Macfarlane2011Oct, wang2015crystallization, he_colloidal_2020, Wang2021May, hensley2022self, hensley2023macroscopic}, fully-addressable structures \cite{Tikhomirov2017MonaLisa, Wintersinger2023Mar, ke_three-dimensional_2012, wei_complex_2012, ong2017programmable}, shells \cite{tikhomirov_triangular_2018, sigl_programmable_2021, bale_accurate_2016}, tubules \cite{shen_novo_2018, Hayakawa2022Oct, videbaek2023economical, Rothemund2004tube}, and sheets \cite{tikhomirov_programmable_2017, Zhang2013May, Tang2023Jun, Rothemund2004Sierpinski}.

One compelling target for self-assembly that has emerged in the past few years is programmable crystalline materials with user-specified unit-cell sizes and symmetries that can be controlled independently of the subunit geometry. In contrast to conventional crystalline materials, in which the lattice parameter is set by the subunit size, such `supercrystals' in principle allow for the precise ordering of molecules or nanoparticles at length scales that can be arbitrarily large in comparison to the building-block dimensions and with symmetries that are decoupled from the subunit shape~\cite{liu_self-organized_2016, Wintersinger2023Mar}. This class of materials is particularly useful for a range of applications, including photonic-plasmonic devices that require the periodic positioning of metallic nanostructures, i.e. `mesoatoms' of complex and precisely defined shapes, at the micrometer length scale to carefully tune plasmonic lattice couplings \cite{Auguie2008Sep,Kravets2018Jun,Bin-Alam2021Feb,Cortes2022Oct}.  However, there is no general method for the optimal design of multi-component assembly to produce a given unit-cell dimension and symmetry. Furthermore, there is no general strategy for how to do so in an economical way, that reaches a given complex 2D crystalline target via the minimal number of unique components and interactions, thereby making the designs easier to implement in practice. 

\begin{figure*}[th]
 \centering
 \includegraphics[width=\linewidth]{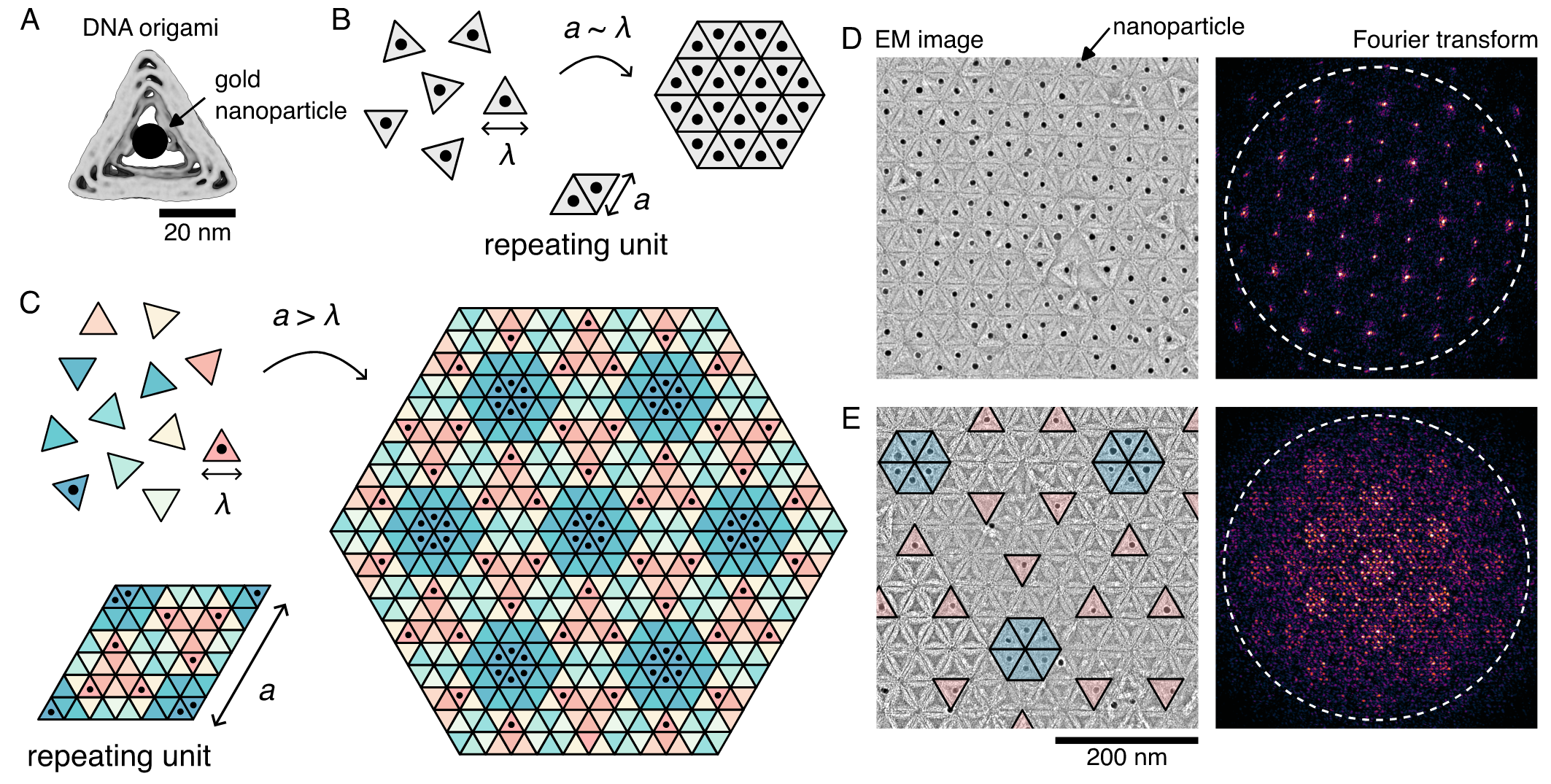}
 \caption{\textbf{Self-assembly of DNA origami triangles into two-dimensional supercrystals.}
(A) Cryo-EM reconstruction of the triangular building block made using DNA origami viewed from the top. DNA-grafted gold nanoparticles (black circle) can be conjugated to the center of the DNA origami using DNA hybridization. (B) A single triangular species assembles into a triangular lattice. Each triangle is illustrated with a dark circle representing a moiety that can be conjugated onto the building block, for example, a gold nanoparticle or an enzyme. The lattice spacing of this system, $a$, roughly matches the size of the building block, $\lambda$. (C) Twelve species of triangles assemble into a tiling with a larger repeating unit. By labeling a subset of the triangles, it is possible to create a supercrystal with a larger lattice parameter and a crystal symmetry that can be different than that of the underlying triangular lattice. (D and E) Experimental demonstrations of supercrystal assembly using (D) one and (E) 12 species of triangles. The designed lattice is that from (B) and (C). In (E), the false colors denote the triangle species that are complementary to gold nanoparticles (black dots). Fourier transforms of the gold nanoparticles show distinct peaks corresponding to the periodicity and the pattern of the gold nanoparticles (see SI Section~VII for image processing details). The dotted circle on the Fourier transform image represents the radius that corresponds to the first zero value of the Bessel function originating from the radius of gold nanoparticles.  
}
 \label{fig:1}
\end{figure*}

Here, we develop a symmetry-based inverse-design method to generate the interaction matrices that specify the assembly of supercrystals of arbitrarily large complexity. We consider equilateral triangular building blocks with programmable interactions on their edges that assemble into two-dimensional (2D) tilings (Fig.~\ref{fig:1}A). In a mixture that contains only one unique component, which we refer to as a species, our subunits crystallize into a simple triangular lattice with a lattice parameter, $a$, that is given by the subunit size, $\lambda$ (Fig.~\ref{fig:1}B). By increasing the number of subunit species, we can assemble more complex tilings with arbitrarily large lattice spacings, $a \gg \lambda$ (Fig.~\ref{fig:1}C). We develop an inverse design method for selecting the most economical designs by exploiting translational and rotational symmetries. To show that our design scheme works in practice, we synthesize DNA origami subunits with programmable interactions (Fig.~\ref{fig:1}A). We assemble tilings with up to 12 species, containing as many as 72 triangles in a unit cell, significantly larger than other recent examples reported in the literature \cite{Kahn2022Sep}. Finally, by conjugating DNA-grafted gold nanoparticles onto a small number of species, we fabricate various gold nanoparticle supercrystals with periodicities reaching hundreds of nanometers (Fig.~\ref{fig:1}D and E). Fourier transforms of the gold nanoparticles exhibit distinct low-spatial-frequency features, demonstrating order on length scales larger than the gold nanoparticles themselves.

\section*{Results}

We consider a system of equilateral triangles that bind to one another through their edges. The triangles have three sides whose interactions are programmed independently (Fig.~\ref{fig:2}A). Note that we limit ourselves to triangles that do not flip in-plane. Under these simple rules, the set of interactions can be represented by a symmetric matrix, in which each element is either filled for favorable interactions or unfilled for unfavorable interactions (Fig.~\ref{fig:2}B).

Perhaps the most naive and straightforward strategy for generating complex 2D patterns for self-assembly is through direct enumeration of all possible interaction matrices. However, this strategy does not work in practice because only a tiny fraction of the interaction matrices encode for unique periodic tilings. For example, there exist only three tilings using a single species, as compared to the $2^6$, or 64, possible interaction matrices (see SI Section~II~A). This problem gets even worse as the number of species, $N$, increases because the number of distinct interaction matrices diverges as $2^{\frac{3}{2}(3N+1)N}$, rendering the enumeration untenable for all but the absolute smallest system sizes. For a more detailed discussion, see SI Section~II.

\subsection{Generating tilings using symmetry}

\begin{figure*}
 \centering
 \includegraphics[width=.99\linewidth]{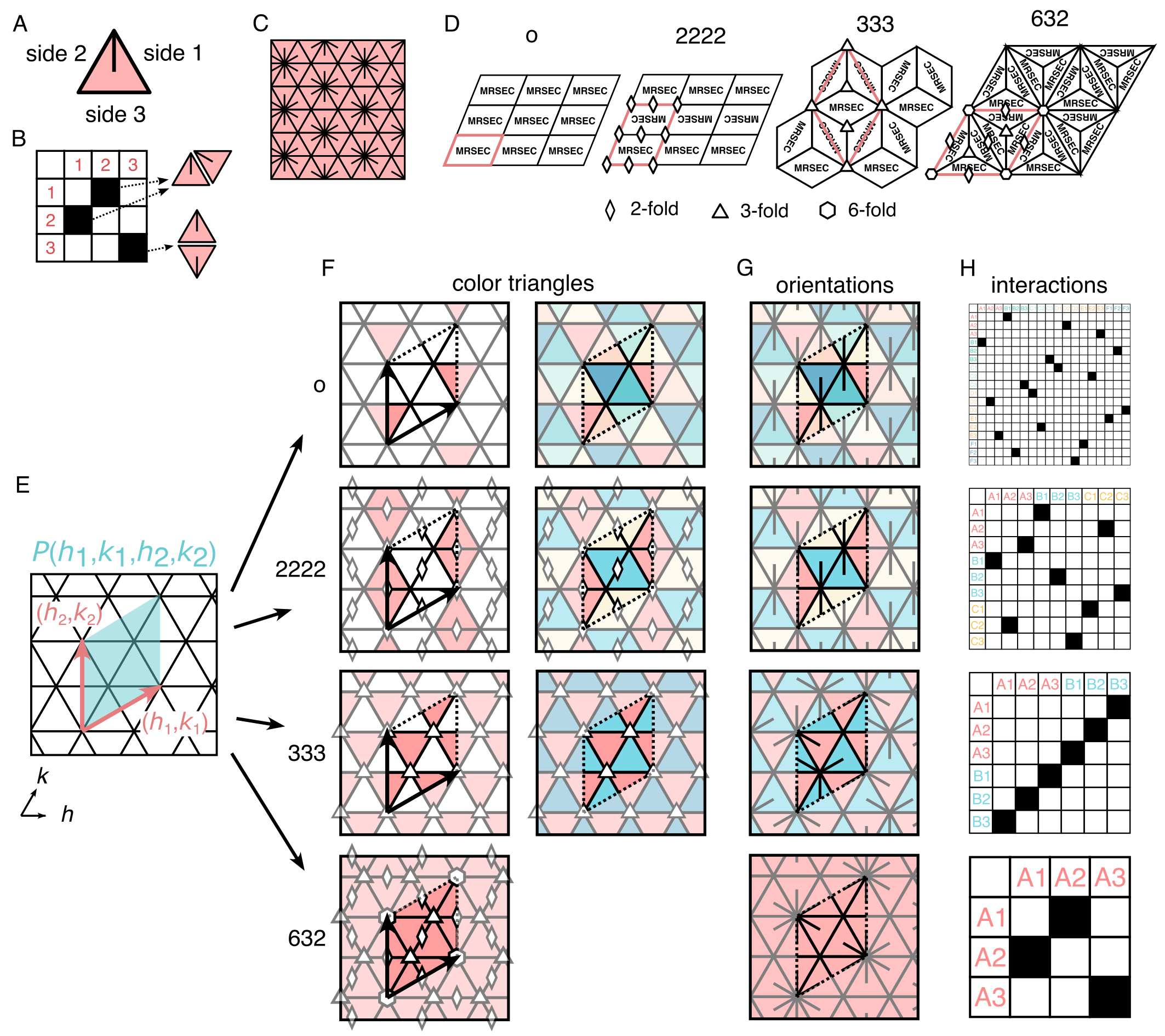}
 \caption{\textbf{Symmetry-based method to generate 2D planar tilings and supercrystals.}
(A) A representative illustration of a triangle. The orientation of the triangle is indicated by a line that points at the vertex between side 1 and 2. Side 1, 2, and 3 always appear counter-clockwise in our schematics. (B) A representative illustration of an interaction matrix for a single species of triangles. Filled squares of the interaction matrices indicate existing interaction between the row and the column of the element. (C) Example of 2D tilings with single species corresponding to the interaction from (B). (D) The four Wallpaper groups that we encounter in our tilings are shown. The parallelogram highlighted in red shows the PU cell, while white symbols indicate the rotational symmetry points. A diamond, triangle, and hexagon symbols represent 2, 3, and 6-fold rotational symmetry points, respectively. The `MRSEC' parallelogram with a black outline indicates the fundamental domain of each tiling. (E) A PU cell, $P(h_1,k_1,h_2,k_2)$, is selected by combining two vectors $(h_1,k_1)$ and $(h_2,k_2)$. The (F) colors and (G) orientations of the individual triangles are chosen such that they obey the imposed symmetry. (H) From the tilings, interaction matrices are inferred. 
}
 \label{fig:2}
\end{figure*}

To circumvent the challenges associated with direct enumeration, we develop a symmetry-based design strategy to generate complex interaction matrices. Each planar tiling can be classified by the symmetry operation that leaves it unchanged. Given their translational symmetry, all 2D planar tilings must fall under one of the 17 Wallpaper groups~\cite{hahn1983international, conway_symmetries_2008}. Here, we use orbifold notation to describe the symmetries of the Wallpaper groups~\cite{Conway2002Aug}. In our system, the allowed symmetry operations are constrained by the interaction rules and the geometry of the particles. Since we prohibit triangles from flipping in-plane, reflections, and glide reflections are not allowed. Additionally, 90-degree rotational symmetries are prohibited in a system with triangular subunits. Therefore, we find only four of the available Wallpaper groups: o tilings with only translational symmetry, 2222 tilings with four distinct 2-fold rotational symmetries, 333 tilings with three distinct 3-fold rotational symmetries, and 632 tilings with 6, 3, and 2-fold rotational symmetries (Fig.~\ref{fig:2}D).

Taking advantage of the Wallpaper symmetries, we develop a tractable approach to designing 2D tilings with arbitrarily large complexity. The core idea is to invert the process of generating tilings: Rather than enumerating every possible interaction matrix and then sifting through them to find those that encode 2D tilings, we directly generate tilings using symmetry and then we infer the interaction matrices that encode the assembly of those tilings. We break down our approach into three steps. 

First, we generate a parallelogram that tiles the plane and corresponds to the periodicity of the resulting tiling (Fig.~\ref{fig:2}E). Specifically, we create a coordinate system $(h,k)$ along two lattice directions with the subunit size being unit length and define a parallelogram using two linearly independent vectors with integer components, $(h_1,k_1)$ and $(h_2,k_2)$. We call these two vectors the primitive unit vectors, and the corresponding parallelogram, the primitive unit cell (PU cell).

Next, we ‘color’ individual triangles and specify their orientations in each parallelogram while enforcing the symmetry operations prescribed by the Wallpaper groups. The procedure consists of: (1) choosing a blank triangle, (2) coloring it with a new color, (3) coloring all symmetrically invariant triangles with the same color, and (4) repeating this process until every triangle has been colored in (Fig.~\ref{fig:2}F). We note that because 2-fold symmetries can be placed on either an edge or a vertex, and 3-fold symmetries can be placed on either a vertex or a face, more than one tiling can exist for a given PU cell for 2222 and 333 symmetries (see SI Section~III). Then we repeat this procedure to specify the orientations of each of the triangles similarly to how we specified the colors (Fig.~\ref{fig:2}G). Here, again, an exception exists for tiles with 3-fold symmetry. When the 3-fold symmetry point lies on the face of a triangle, the three edges of the triangle become homologous, and therefore the triangle does not have a deterministic orientation within the tiling. 

Finally, we derive the interaction matrix that encodes the tiling deterministically by assigning interactions between unique bond pairs (Fig.~\ref{fig:2}H). This procedure amounts to setting the interaction matrix equal to the adjacency matrix. Each unique bond pair is recorded as a filled box in the interaction matrix. 

This design method allows us to generate 2D tilings with a large number of species using a personal computer. To show the feasibility of our approach, we generate an exhaustive list up to PU cells containing $200$ triangles, which consist of 1628 o tilings, 2826 2222 tilings, 52 333 tilings, and 38 632 tilings, totaling 4544 2D tilings. Figures~\ref{fig:4} and S15 show some examples of o, 2222, 333, and 632 tilings. Using a typical PC, computing all the associated interaction matrices takes only a few days, whereas direct enumeration would be impossible (see SI Section~II~B for details) and other modern inverse-design methods, such as SAT-assembly~\cite{romano2020designing,russo2022sat}, start to become intractable above roughly 100 species~\cite{bohlin2023designing}.

\subsection{Self-assembly of 2D supercrystals using DNA origami}

Inspired by the engineering challenge of assembling programmable 2D arrays of plasmonically functional nanostructures~\cite{Kravets2018Jun, Cortes2022Oct}, we demonstrate the use of our symmetry-guided design to template the assembly of supercrystals of 10-nm-diameter gold nanoparticles. To demonstrate the principle, we label the centers of specific subsets of triangles, though more sophisticated multispecies labelings can be achieved by exploiting the addressability of DNA origami \cite{Schnitzbauer2017Jun, Tikhomirov2017MonaLisa, Wintersinger2023Mar}. Given the symmetry rules of the tiling templates, the supercrystals that one can assemble in this way are required to satisfy a small number of constraints. First, the largest possible lattice parameter of the supercrystal is bounded by the PU cell size, $S$. Second, the symmetry of the supercrystal need not be the same as the symmetry of the underlying tiling. And third, the order of symmetry of the supercrystal, $O_\textrm{super}$, cannot be smaller than the order of symmetry of the underlying tiling, $O_\textrm{tiling}$, where the order of symmetry, $O$, is the size of the PU cell divided by the size of the fundamental domain. For example, a 2222 tiling ($O_\textrm{tiling}=2$) can template the assembly of a 632 supercrystal ($O_\textrm{super}=6$), but a 632 tiling ($O_\textrm{tiling}=6$) cannot template the assembly of a 2222 supercrystal ($O_\textrm{super}=2$). 

\begin{figure*}[t]
 \centering
 \includegraphics[width=0.99\linewidth]{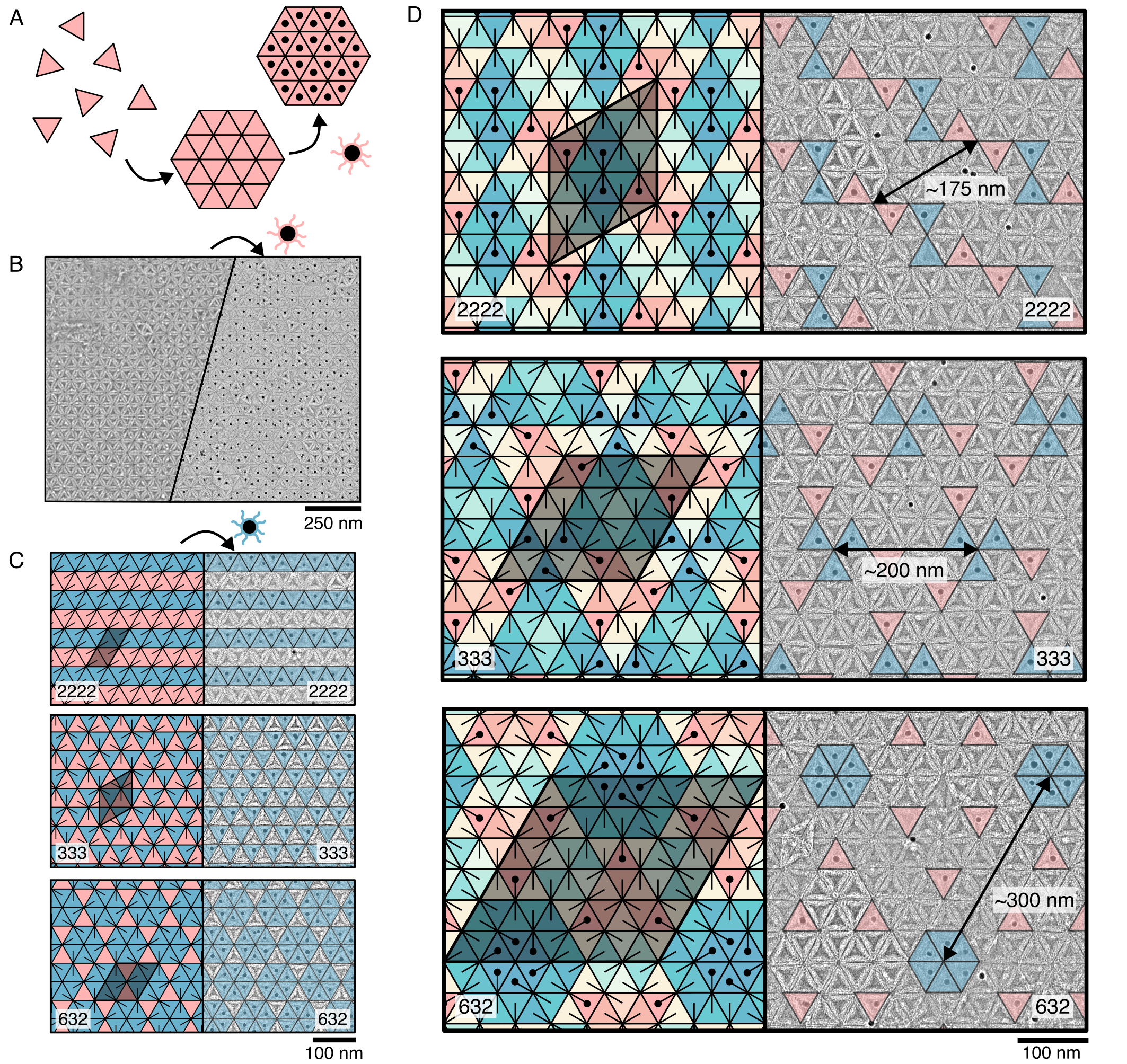}
 \caption{\textbf{DNA origami triangles self-assemble into 2D supercrystals.}
(A) Schematics for assembling supercrystals. To synthesize a supercrystal, triangles are first assembled into tiling, and then labeled using gold nanoparticles. (B) An assembly of flat sheets using a single species of triangles with all homologous, self-complementary edges. TEM micrographs before and after gold nanoparticle labeling are shown. (C and D) 2222, 333, and 632 tilings assembled using (C) 2 and (D) 12 species of triangles. Designed tilings and TEM micrographs are shown side-by-side. The symmetry groups labeled on the left and right indicate the symmetries of the underlying tiling and resulting supercrystal, respectively. Gold-nanoparticle-labeled triangles are highlighted with the species color in the TEM micrograph. The PU cell is indicated by the shaded region. 
}
 \label{fig:4}
\end{figure*}

To demonstrate the power of our design approach, we develop a system based on DNA origami to construct 2D tilings from a large number of distinct species, on which supercrystals of gold nanoparticles can be assembled. Specifically, we make triangular subunits that are roughly 50 nm in edge length and encode specific interparticle interactions using DNA hybridization of sticky ends that protrude from the edges of the subunits (Fig.~\ref{fig:1}A). In this way, we can program the complex interaction matrices that we generate above by exploiting Watson-Crick base pairing. We then assemble DNA origami tilings isothermally at the temperature at which monomers and assemblies coexist, typically around 34~$^\circ$C for the 6-nucleotide sticky ends that we use. Subsequently, we add DNA-grafted gold nanoparticles at a small stoichiometric excess at room temperature to assemble the supercrystals, which we then image using negative-stain transmission electron microscopy (TEM) (Fig.~\ref{fig:4}A--B). We find that the labeling efficiency of gold nanoparticles is roughly 75\% at this stoichiometric ratio.

First, we show that micrometer-sized tilings assemble for the simplest interaction matrix possible. We encode all three sides of a single species of triangle to be homologous and self-complementary. We predict that the triangles in the resulting tiling will have no orientational order, and will correspond to the simplest 632 tiling. Under TEM, we observe 2D sheets spanning micrometers in size---containing over 1,000 subunits---that have the anticipated symmetry (Fig.~\ref{fig:4}B). 

We further demonstrate the utility of our approach by making supercrystals from binary tilings of three different symmetries. By using multiple species with specific interactions, we encode more sophisticated patterns with larger PU cells. First, we assemble representative 2222, 333, and 632 tilings from two species of triangles at the stoichiometric ratios of the tilings (1:1 for 2222 and 333; and 1:3 for 632), as shown as red and blue in Fig.~\ref{fig:4}C, left. The resulting supercrystals, overlayed with false color, are shown in Fig.~\ref{fig:4}C, right. Interestingly, whereas the 2222 and 632 tilings lead to supercrystals with the same symmetry as the parent tiling, the 333 tiling produces a supercrystal with 632 symmetry. This observation highlights the importance of choosing the position of the gold nanoparticle in determining the symmetry of the supercrystal: labeling the center of a triangle yields 632 crystals whereas labeling the edge of a triangle yields 333 crystals, matching the symmetry of the parent.  

Encouraged by the success of our two-species experiments, we assemble more complex supercrystals with larger PU cells, including supercrystals with periodicities comparable to the wavelength of visible light. Figure~\ref{fig:4}D shows examples of 2222, 333, and 632 tilings assembled from 12 species of triangles (see Fig.~S16 for examples of 6-species tilings and SI Section~VI for programmed interactions). We label two triangles for all of the 12-species tilings to maintain an appreciable density of gold nanoparticles. In all cases, we find that the supercrystals are consistent with the underlying tilings, have the same symmetries as their parent tilings, and have low defect densities, indicating that the programmed interactions are truly orthogonal. As anticipated, the size of the PU cell increases as the number of species of triangles increases, leading to larger distances between the gold nanoparticles. This is further seen by the increasing complexity and low-spatial-frequency signal from the Fourier transforms of the gold-particle positions (see SI Section~VII and Fig.~S14 for details). Additionally, we observe that the PU cell size increases with the order of the symmetry of the tiling for a fixed number of species from 175 nm to 200 nm to 300 nm for the 2222, 333, and 632 symmetry tilings, respectively. This final observation hints at the possibility that some tiling patterns might be more useful than others in templating nanoparticle supercrystals.


\begin{figure}[!ht]
 \centering
 \includegraphics[width=\linewidth]{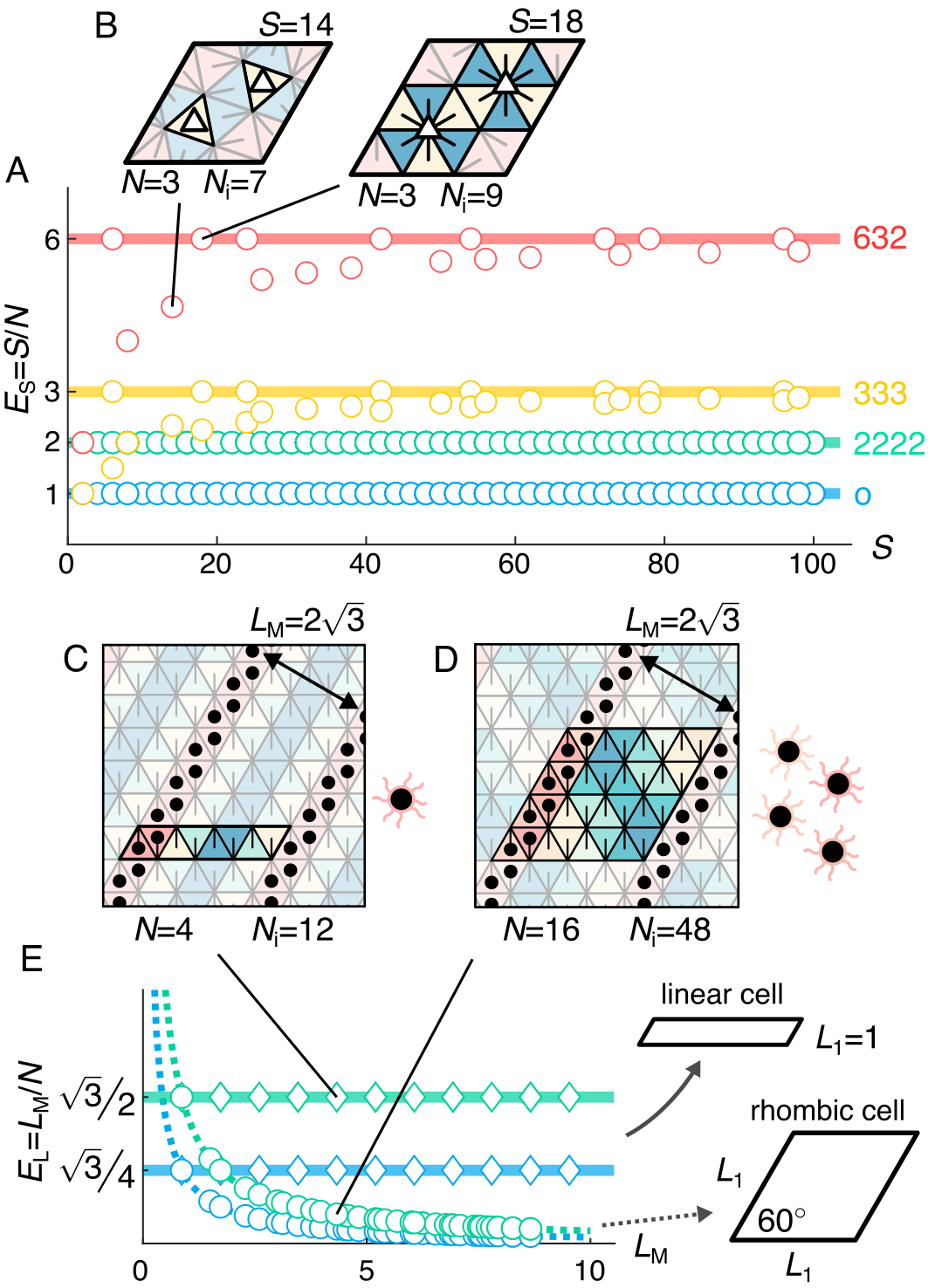}
 \caption{\textbf{Economy of 2D tilings.} 
 (A) The species economy of tiling, $E_\textrm{S}$, for PU cell sizes equal to or smaller than $S=100$ triangles. The upper limit of $E_\textrm{S}$ is $O_\textrm{tiling}$, as illustrated by solid lines, but deviation from this is seen for 333 and 632 tilings. (B) PU cells for two examples of 632 tilings. (B, left) An example 632 tiling that does not satisfy $S/N=O_\textrm{tiling}$. The three sides of the yellow triangles have a homologous interaction since the 3-fold rotational symmetry is located on the triangle face. (B, right) An example 632 tiling that satisfies $S/N=O_\textrm{tiling}$. The 3-fold rotational symmetry is located on the vertex. (C and D) An example of two tilings that can yield the same linear supercrystals with maximum periodic distance $L_\textrm{M}=2\sqrt{3}$. (C) Linear PU cell with $N=4$ species of triangles, one of which is labeled. (D) Rhombic PU cell with $N=16$ species of triangles, four of which are labeled. (E) Linear size economy, $E_\textrm{L}$, is plotted against the maximum periodic distance, $L_\textrm{M}$, for o and 2222 tilings that have linear and rhombic PU cells. Solid lines indicate predicted economy for the linear PU cells whose functional form is given by $E_\textrm{L}=\sqrt{3}O_\textrm{tiling}/4$ while dotted lines indicate rhombic PU cells given by $E_\textrm{L}=3O_\textrm{tiling}/8L_\textrm{M}.$
 }
 \label{fig:5}
\end{figure}

\subsection{Economy of design}

From the perspective of inverse design, we ask the following questions: What is the `cost' of assembling a given tiling, and are some tilings more `economical' than others \cite{duque2023limits}? In the experiment, it is natural to define the cost as the number of species of triangles $N$ that need to be synthesized because each species requires unique DNA staple sequences and must be folded and purified separately. We must also define the `value' of a tiling. Here, we choose the size of the PU cell, $S$, as the value of the tiling, since this parameter controls the spacing of the periodic patterns. Combining these two definitions, the `species economy', $E_\textrm{S}$, of a tiling can be calculated by taking the ratio of the value to the cost, or $E_\textrm{S}=S/N$. 

We find that the species economy has a maximum value that is determined by the order of symmetry of the tilings, $O_\textrm{tiling}$ (Fig.~\ref{fig:5}A). We observe that there is an upper limit to the species economy that can be achieved depending on the symmetry of tilings, which is 1, 2, 3, and 6 for o, 2222, 333, and 632 tilings, respectively. For o and 2222, the species economy for all tilings is constant, while for 333 and 632 tilings, some tilings do not reach the maximum species economy. 

These observations can be rationalized by considering the fundamental domain of the tilings. The size of the fundamental domain is given by $S/O_\textrm{tiling}$, where $O_\textrm{tiling}=1,2,3,6$ for o, 2222, 333, and 632, respectively. As a consequence, the fundamental domain contains all unique triangle species and edges that appear in a tiling. If we simply assume that every unique triangle has to appear in the fundamental domain once, we obtain $S/O_\textrm{tiling}=N$. Combining this expression with our definition of species economy, we obtain $E_\textrm{S}=O_\textrm{tiling}$, which places an upper limit on the economy of any given tiling. 

The loss of species economy in some 333 and 632 tilings can be attributed to triangles having homologous interactions and, therefore, not maximally exploiting all of the information that can be encoded in every subunit. In Fig.~\ref{fig:5}B, we show two examples of 632 tilings, one whose species economy is smaller than the order of symmetry and another whose species economy matches the order of symmetry. The major difference between the two tilings is the location of the 3-fold rotational symmetry point; while the tiling with the higher species economy has a 3-fold symmetry point located at the vertex of the tiling, the one with the smaller species economy has a 3-fold symmetry point located in the middle of a triangle. In the lower species economy case, since the 3-fold rotational symmetry is located in the middle of the triangle, the three sides of that triangle must have the same interactions. Therefore, the interactions encoded in that triangle are redundant, decreasing the overall species economy. In general, whenever there is a 3-fold rotational symmetry placed in the middle of the face of a triangle, the triangle has homologous interactions on three sides and results in a loss of species economy. This feature explains why only 333 and 632 tilings encounter this decrease in the species economy since they are the only tilings in which the symmetry points can have the same symmetry as the 3-fold-symmetric triangular subunits.

Instead of focusing on the number of species of triangles, the `cost' could instead be defined as the number of unique interactions, $N_\textrm{i}$. Indeed, in some systems, the number of unique, orthogonal interactions might be the limiting factor in realizing more and more complex assemblies, such as systems in which the interactions are specified by DNA sequences, magnetic dipoles, or geometric shapes \cite{Wu2012Nov,gerling_dynamic_2015,Huntley2016May,Niu2019Dec}.
Surprisingly, within this definition of economy, which we refer to as the interaction economy, $E_\textrm{i}=S/N_\textrm{i}$, every planar tiling discussed in this paper follows a simple equation: 
\begin{equation}
    E_\textrm{i}= O_\textrm{tiling}/3,
\end{equation}
irrespective of the specific locations of the symmetry points.
This relationship can be understood as a consequence of the fact that the number of unique edges contained in the fundamental domain is always $N_\textrm{i} = 3S/O_\textrm{tiling}$. We note that the number of unique species and interactions are just two potential metrics of the `cost', and other strategies for quantifying the information content of complex assemblies might lead to other definitions of economy \cite{Ahnert2010Aug}.

\subsection{Economical design of linear supercrystals}

We conclude by considering a specific class of supercrystals, in which the periodic distance between rows of nanoparticles is the only relevant design constraint. Figures~\ref{fig:5}C and D show two example tilings that yield the same supercrystal, with nanoparticles labeled in lines that are periodic in one of the other two lattice directions. In one case, the tiling requires four unique triangles, one of which is labeled to construct the supercrystal (Fig.~\ref{fig:5}C). In the other case, it takes 16 unique triangles, of which four are labeled (Fig.~\ref{fig:5}D). Because we are only concerned with the maximum periodic distance, $L_\textrm{M}$, defined as the length of the PU cell perpendicular to the short axis, we define another economy, linear size economy, as the number of unique particles it takes to program the maximum periodic distance, or $E_\textrm{L}=L_\textrm{M}/N$. 

We find that the linear size economy of supercrystals depends on a combination of the symmetry group and the aspect ratio of the PU cell, with highly anisotropic PU cells of high order being the most economical. Figure~\ref{fig:5}E shows the maximum periodic distance of o and 2222 tilings that have linear and rhombic PU cell shapes. Here, we define a linear PU cell to have length $L_1=1$ and a rhombic PU cell to have $L_1=L_2$, where $L_1$ and $L_2$ are the lengths of short and long edges of the PU cell, respectively. For o and 2222 tilings, the values of $L_\textrm{M}$ can be predicted using a simple geometrical argument. Specifically, for 2222 tilings, the periodicity is given by $L_\textrm{M}=\sqrt{3}N/2$ for linear PU cells and ${L_\textrm{M}}^2=3N/4$ for rhombic PU cells. As before, o tilings are less economical, as the PU cell is half the size of the 2222 PU cell for the same number of subunit species. These results show the importance of the shape of the PU cell for the linear size economy: Tilings with linear PU cells are more economical than rhombic ones. 

\section*{Discussion and Conclusions}

In summary, we developed an inverse design method to create self-assembling tilings and supercrystals of arbitrary size and complexity by exploiting their underlying symmetries. By identifying the right symmetry tilings and the right aspect ratio of the primitive unit cell, our strategy is guaranteed to yield economical, deterministic designs. Here, the two central take-home messages are: (1) Tilings of the highest order symmetry, 632 in our case, yield the largest PU cell for the smallest number of components; and (2) Highly anisotropic PU cells with 2222 symmetry yield the most economical linear supercrystals. Using DNA origami, we demonstrated how our strategy could be used to assemble supercrystalline arrangements of 10-nm-diameter gold nanoparticles with periodicities up to 300 nm, comparable to the wavelength of visible light. We are optimistic that this work will help to pave the way toward making user-defined tilings and crystals of arbitrary complexity, ushering in the possibility of making mesoscopic materials with complex plasmonic and photonic functionality via self-assembly.

Whereas in the main text, our initial definition of economy focused on the economic cost of synthesizing many subunit species, we highlight another cost associated with increasing the assembly complexity: the time to assembly. Specifically, we found that increasing the number of species in a tiling requires increasing the time to assembly. Whereas single-species systems assembled micrometer-sized tilings in less than a day, similarly sized tilings made from 12 species took roughly two weeks to assemble at the same temperature (see Fig.~S17). This observation indicates that the growth rate decreases roughly linearly with the number of species, which is consistent with detailed measurements of the growth rate of multi-species tubules using a similar experimental system \cite{videbaek2023economical}. Although not explored in this paper, we anticipate that hierarchical or nonequilibrium approaches  \cite{Tikhomirov2017MonaLisa, Jacobs2015periodic, Wintersinger2023Mar, Gartner2022Jan} may help to overcome these kinetic `costs' of multi-species assembly.

Though the general principle of economy introduced in this paper can easily be extended to other conventional tilings and crystal systems, naturally, one may wonder how the system compares to nonconventional ones, such as the crisscross structure \cite{Minev2021Mar,Wintersinger2023Mar}. Unlike a typical tiling, where a polygon interacts with other polygons through their edges, the component of crisscross structures are slats which can interact with other slats at designated points along the slat length. A unique feature of the crisscross system is that the number of interaction sites per component is much larger than that achievable in typical tiling systems, allowing for a robust design of the nucleation barrier. To compare the economy of the slate system with that of tilings, we introduce a new metric of economy. Whereas the `cost' of assembly can be defined as the number of particle species as before, we define the new `value' as the area of the PU cell normalized by the representative subunit length. Defining economy as the value over the cost, we observe that the economy of the triangular tiling is much larger than that of the crisscross structures (for details, see SI Section~V). This result agrees with the intuition that crisscross systems have increased connectivity between components spanning smaller areas, which leads to a limited economy. Therefore, though the crisscross system is advantageous regarding the suppression of spurious nucleation, it may require more unique components as compared to conventional tiling designs.

Finally, we conclude by highlighting that our symmetry-based inverse design method can be extended to other tilings and is not constrained by the shape of the building block, including both 2D tilings and 3D crystals. In this work, we specifically focused on orientable triangle tilings, in which the corresponding symmetry groups include o, 2222, 333, and 632. However, depending on the specific target structure or application, one may prefer to fabricate 2D supercrystals with different symmetries or even 3D supercrystals. We emphasize that the procedures developed herein can be universally applied to other systems that have translational symmetry. In SI Section~IV, we summarize general inverse design strategies that can be applied to translationally symmetric structures and show examples, including square, hexagonal, parallelogram, rhombille, and snub-square tilings. Therefore, the inverse design method we developed can be applied to self-assemble other 2D and 3D supercrystals of arbitrary complexity. In addition, we speculate that the design rules we identified have connections to 2D manifolds, which opens up new design spaces that can be accessed by controlling the dihedral angles between neighboring subunits to create curved structures \cite{caspar_physical_1962,tanaka2023programmable, videbaek2023economical, sigl_programmable_2021, duque2023limits}.

\begin{acknowledgments}
TEM images were prepared and imaged at the Brandeis Electron Microscopy facility. This work is supported by the Brandeis University Materials Research Science and Engineering Center, which is funded by the National Science Foundation under award number DMR-2011846. D.H. acknowledges support from the Masason Foundation. We acknowledge Oleg Gang for guidance on conjugating DNA to gold nanoparticles and Ian Murphy for reducing thiolated DNA.
\end{acknowledgments}

\bibliography{main.bib}

\end{document}


\title{Supplemental Information for ``Symmetry-guided inverse design of self-assembling multiscale DNA origami tilings"}
\author{Daichi Hayakawa, Thomas E. Videb\ae k, Gregory M. Grason, W. Benjamin Rogers}

\maketitle

\section{Experimental methods}\label{sec:methods}

\subsection{Folding DNA origami}\label{subsec:folding}
Each DNA origami particle is folded by mixing 50~nM of p8064 scaffold DNA (Tilibit) and 200~nM each of staple strands with folding buffer and annealed through a temperature ramp starting at 65~$^{\circ}$C for 15 minutes, then 58 to 50~$^{\circ}$C, $-1~^{\circ}$C per hour. Our folding buffer, contains 5~mM Tris Base, 1~mM EDTA, 5~mM NaCl, and 15~mM MgCl$_2$. We use a Tetrad (Bio-Rad) thermocycler for annealing the solutions. 

\subsection{Agarose gel electrophoresis}\label{subsec:electrophoresis}
To assess the outcome of folding, we separate the folding mixture using agarose gel electrophoresis. Gel electrophoresis requires the preparation of the gel and the buffer. The gel is prepared by heating a solution of 1.5\% w/w agarose, 0.5x TBE to boiling in a microwave. The solution is cooled to 60~$^{\circ}$C. At this point, we add MgCl$_2$ solution and SYBR-safe (Invitrogen) to adjust the concentration of the gel to 5.5~mM MgCl$_2$ and 0.5x SYBR-safe. The solution is then quickly cast into an Owl B2 gel cast, and further cooled to room temperature. The buffer solution contains 0.5x TBE and 5.5~mM MgCl$_2$, and is chilled to 4~$^{\circ}$C before use. Agarose gel electrophoresis is performed at 110 V for 1.5 to 2 hours in a cold room kept at 4~$^{\circ}$C. The gel is then scanned with a Typhoon FLA 9500 laser scanner (GE Healthcare).

\subsection{Gel purification and resuspension}\label{subsec:purification}
After folding, DNA origami particles are purified to remove all excess staples and misfolded aggregates using gel purification. The folded particles are run through an agarose gel (now at a 1xSYBR-safe concentration for visualization) using a custom gel comb, which can hold around 4~mL of solution per gel. We use a blue fluorescent table to identify the gel band containing the monomers. The monomer band is then extracted using a razor blade, which is further crushed into smaller pieces by passing through a syringe. We place the gel pieces into a Freeze `N' Squeeze spin column (Bio-Rad), freeze it in a -80~$^\circ$C freezer for 30 minutes, thaw at room temperature, and then spin the solution down for 5 minutes at 13 krcf. 

Since the concentration of particles obtained after gel purification is typically not high enough for assembly, we concentrate the solution through ultrafiltration~\cite{wagenbauer_how_2017}. First, a 0.5~mL Amicon 100kDA ultrafiltration spin column is equilibrated by centrifuging down 0.5~mL of the folding buffer at 5~krcf for 7 minutes. Then, the DNA origami solution is added up to 0.5~mL and centrifuged at 14~krcf for 15 minutes. Finally, we flip the filter upside down into a new Amicon tube and spin down the solution at 1~krcf for 2 minutes. The concentration of the DNA origami particles is measured using a Nanodrop (Thermofisher), assuming that the solution consists only of monomers, where each monomer has 8064 base pairs.

\subsection{Tile assembly}\label{subsec:assembly}
All assembly experiments are conducted at a DNA origami particle concentration of 10~nM. For multispecies tilings, the total DNA origami concentration is 10~nM and each triangular species are mixed in a stoichiometric ratio of the target tilings. By mixing the concentrated DNA origami solution after purification with buffer solution, we make 50~$\upmu$L of 10~nM DNA origami at 20~mM MgCl$_2$. The solution is carefully pipetted into 0.2~mL strip tubes (Thermo Scientific) and annealed through different temperature protocols using a Tetrad (Bio-Rad) thermocycler.

\subsection{Negative stain TEM}\label{subsec:TEM}
We first prepare a solution of uranyl formate (UFo). ddH$_2$O is boiled to deoxygenate it and then mixed with uranyl formate powder to create a 2\% w/w UFo solution. The solution is covered with aluminum foil to avoid light exposure, then vortexed vigorously for 20 minutes. The solution is filtered using a 0.2~$\upmu$m filter. The solution is divided into 0.2~mL aliquots, which are stored in a --80~$^\circ$C freezer until further use.

Prior to each negative-stain TEM experiment, a 0.2~mL aliquot is taken out from the freezer to thaw at room temperature. We add 4~$\upmu$L of 1~M NaOH and vortex the solution vigorously for 15 seconds. The solution is centrifuged at 4~$^\circ$C and 16~krcf for 8 minutes. We extract 170~$\upmu$L of the supernatant for staining and discard the rest. 

The EM samples are prepared using FCF400-Cu grids (Electron Microscopy Sciences). We glow discharge the grid prior to use at --20~mA for 30 seconds at 0.1~mbar, using a Quorum Emitech K100X glow discharger. We place 4~$\upmu$L of the sample on the grid for 1 minute to allow adsorption of the sample to the grid. During this time 5~$\upmu$L and 18~$\upmu$L droplets of UFo solution are placed on a piece of parafilm. After the adsorption period, the remaining sample solution is blotted on a Whatman filter paper. We then touch the carbon side of the grid to the 5~$\upmu$L drop and blot it away immediately to wash away any buffer solution from the grid. This step is followed by picking up the 18~$\upmu$L UFo drop onto the carbon side of the grid and letting it rest for 30 seconds to deposit the stain. The UFo solution is then blotted to remove excess fluid. Grids are allowed to dry for a minimum of 15 minutes before insertion into the TEM.

We image the grids using an FEI Morgagni TEM operated at~80 kV with a Nanosprint5 CMOS camera (AMT). Images are acquired between x8,000 to x28,000 magnification. The images are high-pass filtered and the contrast is adjusted using Fiji~\cite{Schindelin2012Jul}. 

\subsection{Gold nanoparticle conjugation to DNA origami}\label{subsec:gnp}

We first attach thiol-modified ssDNA (5'-HS-C$_6$H$_{12}$-TTTTTAACCATTCTCTTCCT-3', IDT) to 10~nm diameter gold nanoparticles (Ted Pella) using a protocol similar to that in ref.~\cite{Sun2020valence}. Thiolated strands are first reduced using tris(2-carboxyethyl) phosphine (TCEP) solution (Sigma-Aldrich). 20~mM TCEP (pH 8) and 100~$\upmu$M thiol-DNA are held at room temperature for one hour on a vortex shaker. Excess TCEP is removed with a 10kDa Amicon filter in three washes of a 50~mM HEPES buffer (pH 7.4); filter centrifugation is done at 4~krcf for 50 minutes at 4~$^\circ$C. After purification, thiolated DNA strands are stored at -20~$^\circ$C until needed. To attach thiolated DNA to gold nanoparticles they are mixed at a ratio of 300:1 in a 1x borate buffer (Thermo Scientific) and are rotated at room temperature for 2~hours. After incubation, the salt concentration is increased in a stepwise manner to 0.25~M NaCl using a 2.5~M NaCl solution in five steps. After each salt addition, the gold nanoparticle solution is rotated at room temperature for 30~minutes. After the last addition, the gold nanoparticle solution is left to age in the rotator overnight. To remove excess thiol-DNA strands, DNA-gold nanoparticle conjugates were washed four times by centrifugation using a 1x borate buffer with 0.1~M NaCl. DNA-gold nanoparticle solutions were centrifuged at 6.6~krcf for 1~hour for each wash step. After the last wash, the DNA-gold nanoparticle concentration was measured using a Nanodrop and the solution was stored at 4~$^\circ$C

To attach gold nanoparticles to the tilings we incorporate handles with a complementary sequence to the label on the gold nanoparticle (5'-AGGAAGAGAATGGTT-3', IDT) on the interior edges of the DNA origami subunit. For 2- and 6-species assembly, only one subunit type has handles that bind to the gold nanoparticles and for 12-species assembly, two subunit types have handles that bind to the gold nanoparticles. After the tilings have been assembled, the assembly solution is diluted into a mixture with gold nanoparticles and incubated at 30~$^\circ$C overnight in a buffer containing 5~mM Tris Base, 1~mM EDTA, 5~mM NaCl, and 20~mM MgCl$_2$. The concentrations of samples used for TEM micrographs in this paper are shown in Table~\ref{tab:aunpConc}. After incubation, the samples are prepared for imaging.

\begin{table}[ht]
    \centering
    \begin{tabular}{c|c|c|c|c}
        Sample & Total monomer conc. & Labelable monomer conc. & AuNP conc. & AuNP to labelable monomer ratio\\
        \hline
        632 N=1 & 1 & 1 & 5 & 5\\
        2222 N=2 & 3 & 1.5 & 8 & 5.33\\
        333 N=2 & 1 & 0.5 & 2.5 & 5\\
        632 N=2 & 1 & 0.75 & 3.75 & 5\\
        2222 N=6 & 3 & 0.5 & 6 & 12\\
        333 N=6 & 3 & 0.5 & 6 & 12\\
        632 N=6 & 3 & 0.56 & 6 & 10.67\\
        2222 N=12 & 3 & 0.5 & 5 & 10\\
        333 N=12 & 3 & 0.56 & 3 & 5.33\\
        632 N=12 & 3 & 0.5 & 3 & 6\\
    \end{tabular}
    \caption{\textbf{Concentrations of DNA origami monomers and gold nanoparticles used for conjugation.} All concentrations are in the units of nanomolar.}
    \label{tab:aunpConc}
\end{table}

\subsection{Cryo-electron microscopy}\label{subsec:cryo}
Higher concentrations of DNA origami are used for cryo-EM grids than for assembly experiments. To avoid assembly and aggregation of the subunits, we removed ssDNA strands protruding from the faces of the DNA origami. To prepare samples we fold 2~mL of the folding mixture, gel purify it, and concentrate the sample by ultrafiltration, as described above, targeting a concentration of 300~nM of DNA origami. EM samples are prepared on glow-discharged C-flat 1.2/1.3 400 mesh grids (Protochip). Plunge-freezing of grids in liquid ethane is performed with an FEI Vitrobot with sample volumes of 3~$\upmu$L, blot times of 16~s, a blot force of -1, and a drain time of 0 seconds at 20~$^\circ$C and 100\% humidity.


Cryo-EM images for the DNA origami monomer were acquired with the Tecnai F30 TEM with a field emission gun electron source operated at 300~kV and equipped with an FEI Falcon II direct electron detector at a magnification of x39000. Single particle acquisition is performed with SerialEM. The defocus is kept at --2~$\upmu$m with a pixel size of 2.87 Angstrom.

\subsection{Single-particle reconstruction}\label{subsec:recontruction}
\begin{figure}[t]
 \centering
 \includegraphics[width=0.9\linewidth]{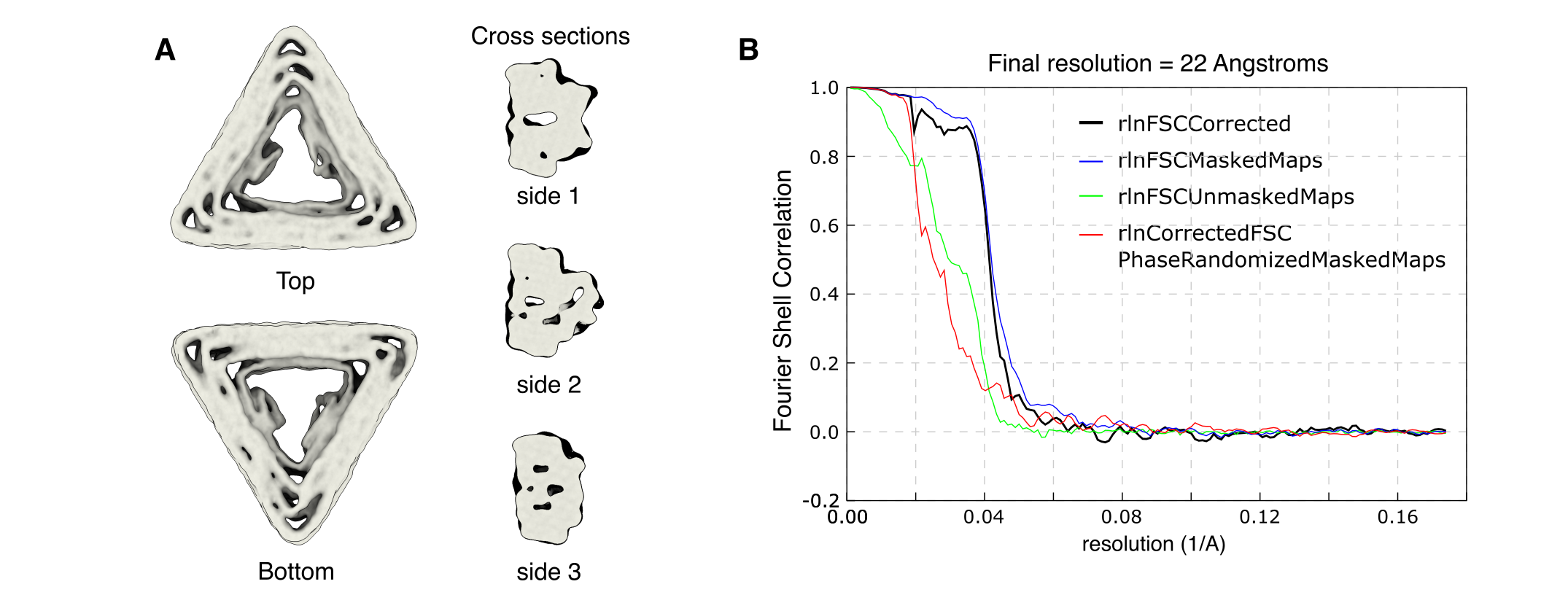}
 \caption{\textbf{Cryo EM reconstruction of the DNA origami particles} (A) Cryo EM reconstruction of the DNA origami particles along with its cross-sectional views. (B) Plot of the Fourier shell correlation curves used to estimate the resolution of the DNA origami particle.}
 \label{fig:cryo}
\end{figure}

Image processing is performed using RELION-3~\cite{Zivanov2018RELION3}. Contrast-transfer-function (CTF) estimation is performed using CTFFIND4.1~\cite{ctffind}. After picking single particles we performed a reference-free 2D classification from which the best 2D class averages are selected for processing, estimated by visual inspection. The particles in these 2D class averages are used to calculate an initial 3D model. A single round of 3D classification is used to remove heterogeneous monomers and the remaining particles are used for 3D auto-refinement and post-processing.  Our reconstruction for the monomer uses 2650 particles and has a resolution of 21.3~\r{A}, Fig.~\ref{fig:cryo}. The post-processed map is deposited in the Electron Microscopy Data Bank.


\section{Patterns generated by enumerating interaction matrices}\label{sec:enum}

\subsection{Interaction matrix generation for single species of triangles}\label{sec:single}

\begin{figure}[ht]
 \centering
 \includegraphics[width=0.9\textwidth]{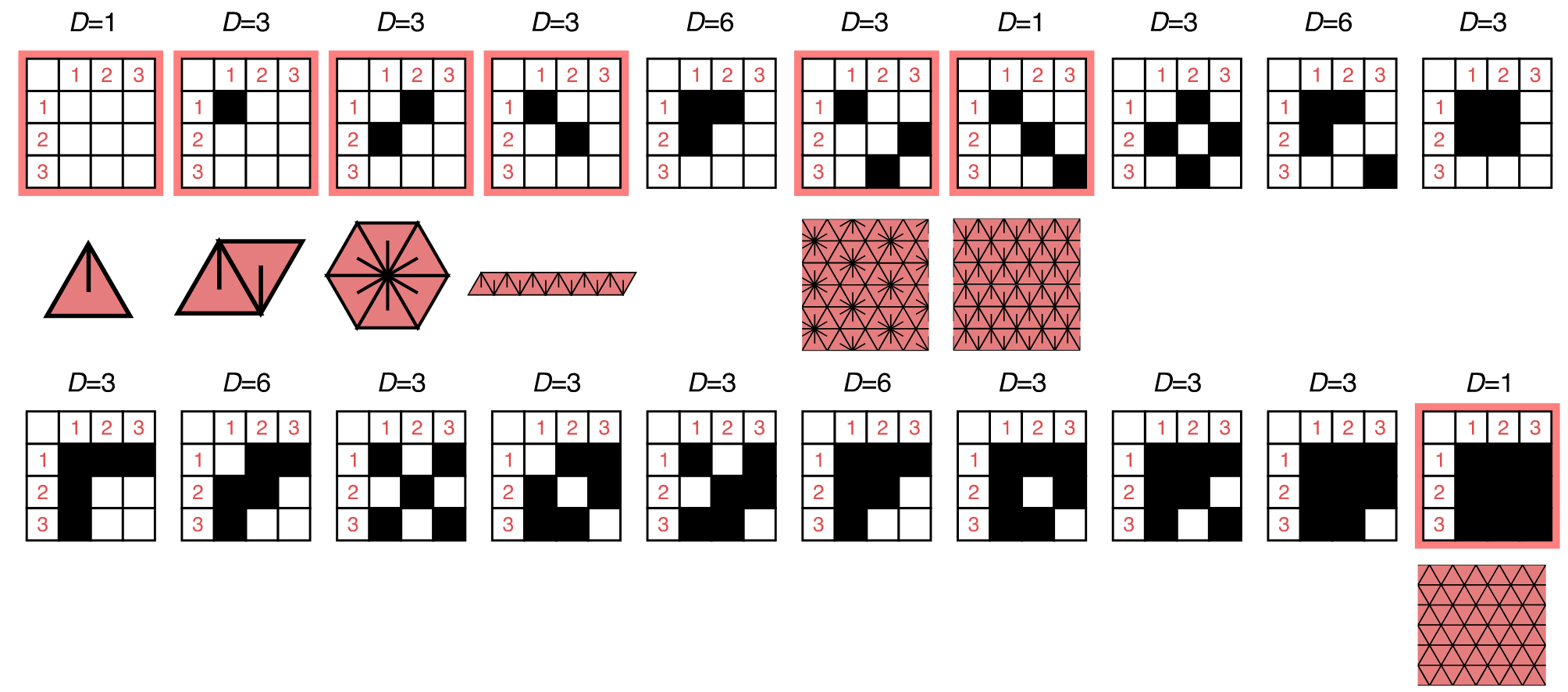}
 \caption{\textbf{Enumerating interaction matrices for single species of triangle.} $D$ denotes the degeneracy of the interaction matrix through rotation or reflection. Interaction matrices highlighted in red are fully deterministic patterns and are accompanied by corresponding illustrations of the pattern.}
 \label{Sfig:monoMatrix}
\end{figure}

For a single species of triangles, we illustrate the 24 unique interaction matrices in Fig.~\ref{Sfig:monoMatrix}. The reduction from the original $2^6$ or 64 interaction matrices comes from the degeneracy in labeling the sides with numbers. For example, interaction matrices that match by shifting the side numbers circularly are rotationally degenerate, thus reducing the number of unique interaction matrices. Specifically, the degeneracy of an interaction matrix is determined by how many different interaction matrices it can represent through cyclic rotations of side 1, 2, and 3, swapping of two sides corresponding to reflection, or a combination of both. The degree of degeneracy of each matrix varies between one, three, and six, depending on the configurations.  For example, the second interaction matrix in Fig.~\ref{Sfig:monoMatrix} has a degeneracy of three, since cyclic rotation or swapping can take this interaction matrix from having side 1-1 binding to 2-2 or 3-3 binding. The sum of all degeneracies adds up to 64, which is the number of all possible interaction matrices. 

However, many of these interaction matrices do not have a unique assembly structure corresponding to them. Seven interaction matrices highlighted in red in Fig.~\ref{Sfig:monoMatrix} correspond to fully deterministic patterns, each having a single ground-state assembly structure where all programmed bonds are satisfied (for detailed definition, see Supplementary Section \ref{sec:deterministic}). All others correspond to nondeterministic patterns, or systems without a unique ground-state assembly structure. For such systems, the orientation of triangles being added onto an existing assembly is sometimes ambiguous, leading to various assembly outcomes.

We classify the deterministic tilings into self-limiting clusters, linear tilings, and 2D planar tilings. Self-limiting clusters are finite-sized structures, linear tilings are infinite structures extending in 1D, and 2D planar tilings span 2D with infinite building blocks. It is straightforward to show that deterministic linear and planar tilings are translationally symmetric in 1D and 2D, respectively. In an infinite tiling with a single species of triangle, one can always find a pair of triangles with the same orientation. Since the tiling is deterministic, the assembly procedure between these two pairs of triangles is repeated, yielding an infinite number of triangles with the same orientation. Similarly, deterministic linear and planar tilings with a finite number of species of triangles have translational symmetry. For a single species system, we find three clusters, one linear tiling, and three planar tilings (Fig.~\ref{Sfig:nondet}).

\subsection{Interaction matrix generation for multiple species of triangles}\label{sec:multi}

\begin{figure}[ht]
 \centering
 \includegraphics[width=0.99\textwidth]{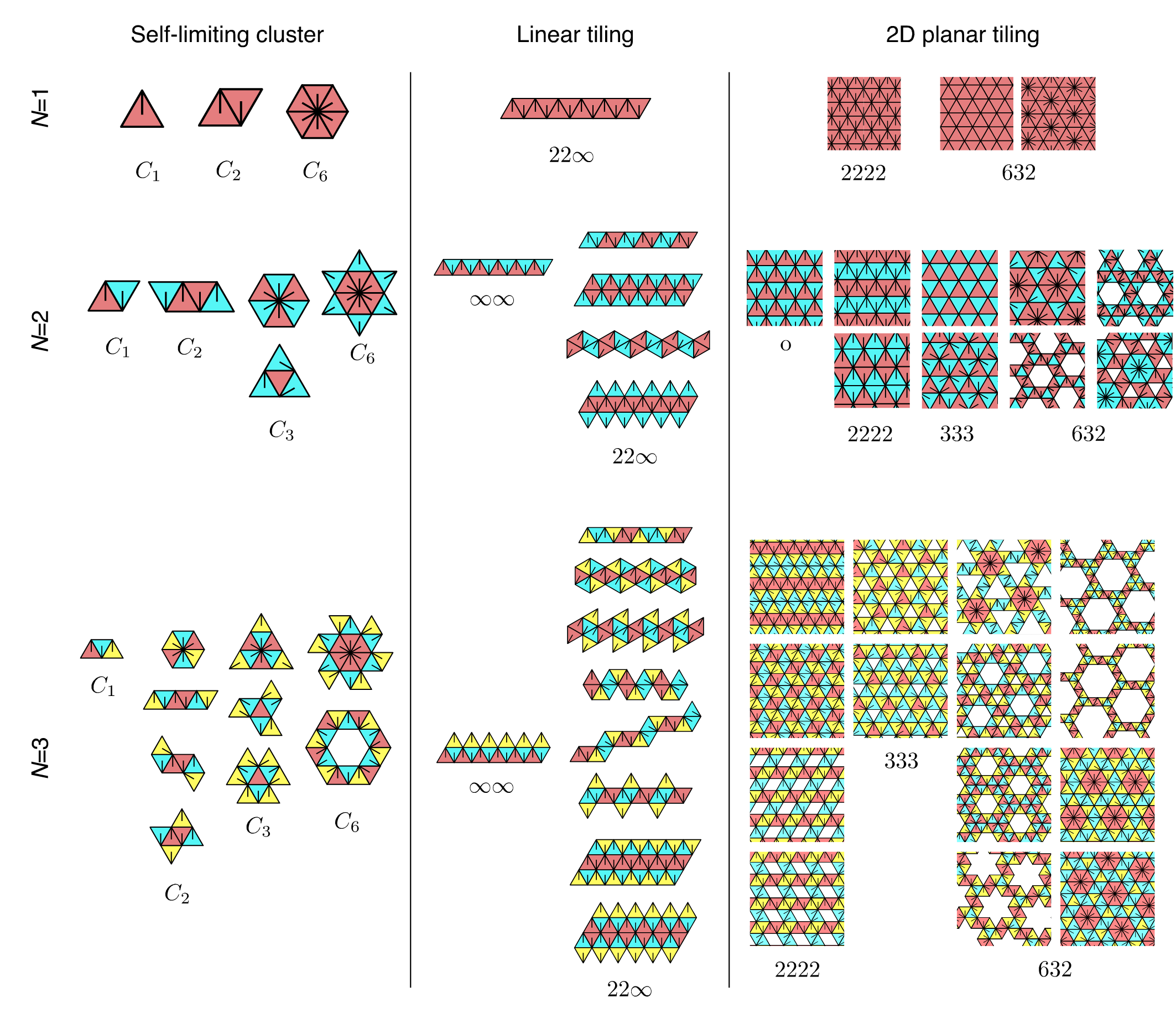}
 \caption{\textbf{Fully deterministic patterns generated by enumerating interaction matrices for up to three species of triangles.} The deterministic patterns for N=1, 2, and 3 species of triangles are separated into clusters, linear tilings, and planar tilings. The notation indicates the symmetry of the patterns.}
 \label{Sfig:detTile}
\end{figure}

Next, we extend our pattern generation by direct enumeration to a system with multiple species of triangles. For two species of triangles, the size of the interaction matrix increases to six-by-six, with $2^{21}$ or 2,097,152 unique combinations. In contrast to a large pool of possible interaction matrices, we only find 19 deterministic patterns, consisting of five self-limiting clusters, five linear tilings, and nine 2D planar tilings (Fig.~\ref{Sfig:detTile}). Within the 2D planar tilings, we find that three tilings have vacancies placed periodically. Similarly, for three species, we find ten self-limiting clusters, nine linear tilings, four filled planar tilings, and ten holey planar tilings among $2^{45}$ possible interaction matrices (Fig.~\ref{Sfig:detTile}). 

\begin{figure}[ht]
 \centering
 \includegraphics[width=0.9\textwidth]{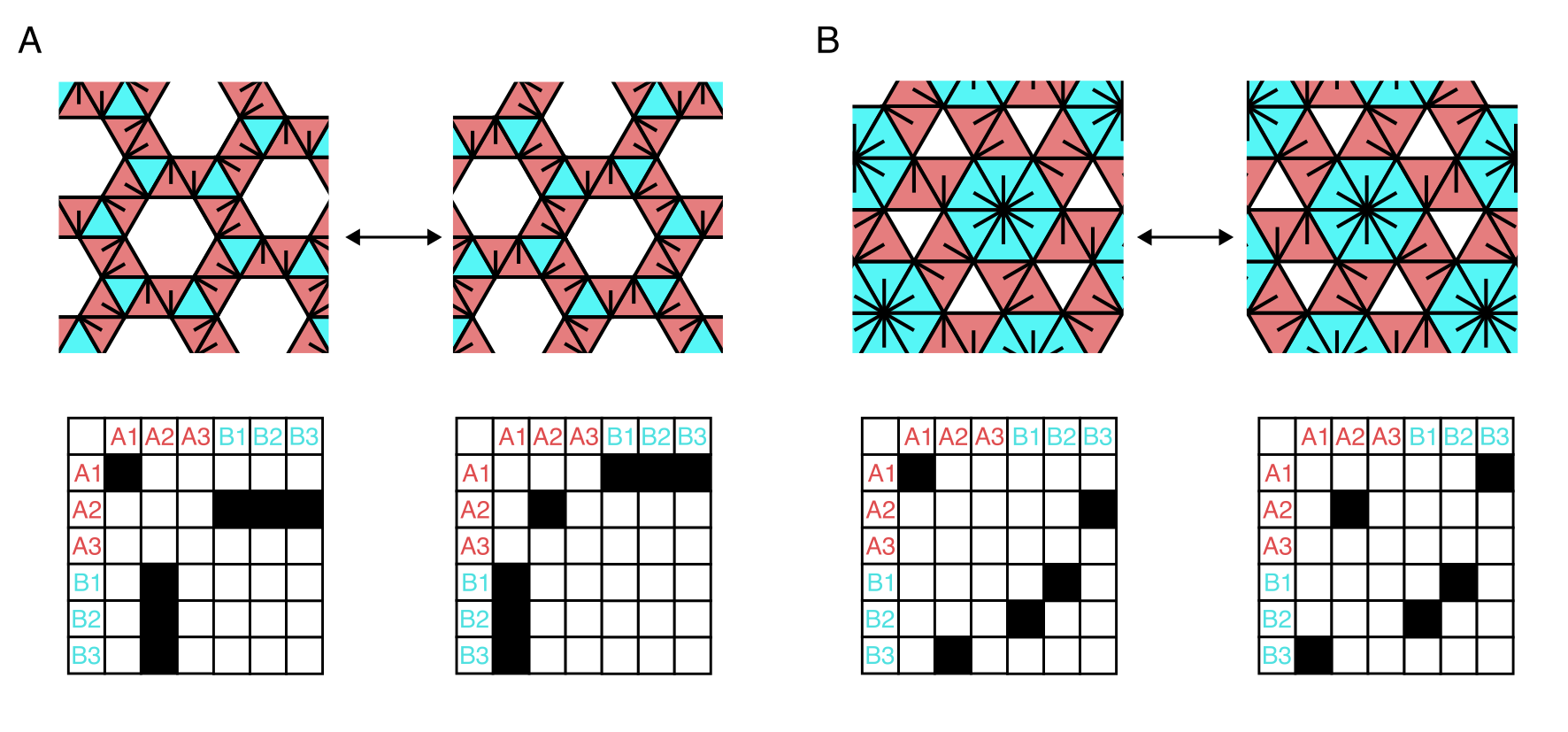}
 \caption{\textbf{Chiral tilings.} Two examples of deterministic chiral tilings and their enantiomorphs. The interaction matrices for all the tilings are also illustrated.}
 \label{Sfig:chirality}
\end{figure}

In the counting of patterns and tilings, we ignore chiral counterparts. Some patterns we generate are enantiomorphic; though the two patterns cannot be matched only through rotations, a combination of rotations and reflections allows the transformation of one into the other (Fig.~\ref{Sfig:chirality}). In this paper, we consider the chiral pattern and its enantiomorph to be the same pattern and are counted as a single pattern. This is because an enantiomorph can be generated simply by reflecting every single component in the system. For example, Fig.~\ref{Sfig:chirality} shows two chiral tilings whose enantiomorphs are generated by flipping the interactions of sides 1 and 2 for both triangles. 

All fully deterministic patterns -- including clusters and linear tilings -- can be classified using symmetry. Whereas we used Wallpaper groups for planar tilings, clusters can be classified using 2D point groups and linear tilings, using Frieze groups~\cite{conway_symmetries_2008}. Specifically, clusters can be classified into cyclic groups with 1, 2, 3, or 6-fold rotational symmetry in the middle of the cluster, which are denoted as $C_1$, $C_2$, $C_3$, and $C_6$ in Fig.~\ref{Sfig:detTile}. Similarly, linear tilings that are generated here can be classified into two of the Frieze patterns, $\infty\infty$ or $22\infty$; $\infty\infty$ is a pattern with only translational symmetry in one direction, whereas $22\infty$ has a 2-fold rotational symmetry. The order of symmetry for the patterns is 1, 2, 3, and 6 for $C_1$, $C_2$, $C_3$, and $C_6$, respectively, for point symmetry groups and 1 and 2 for $\infty\infty$ and $22\infty$ for Frieze groups, respectively. Given the order of symmetry, the number of unique, interacting edges required to assemble a pattern is 
\begin{equation}
    N_i = \frac{3S-E_\mathrm{free}}{O},
    \label{seq:numedge}
\end{equation}
where $E_\mathrm{free}$ is the number of unbound edges in the PU cell and $S$ is the total number of triangles in the PU cell, excluding the holey region. For clusters, we define $S$ to be the entire cluster, since there are no translational symmetry. The equation also describes the number of unique edges necessary for holey 2D planar tilings. Equation~(\ref{seq:numedge}) applies only to fully-deterministic tilings and it cannot be extended to globally deterministic tilings, which are defined in Supplementary Section \ref{sec:deterministic}. 

\subsection{Definition of deterministic and nondeterministic tilings}\label{sec:deterministic}

\begin{figure*}[ht]
 \centering
 \includegraphics[width=0.9\textwidth]{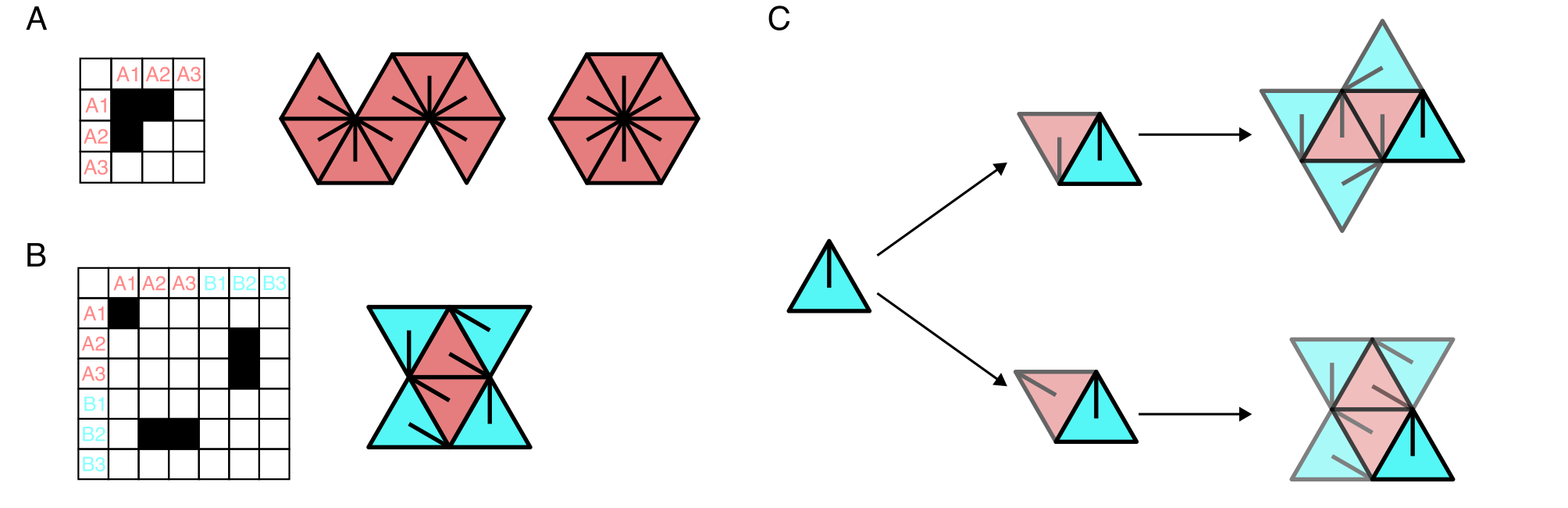}
 \caption{\textbf{Examples of systems with nondeterministic and deterministic interactions.} (A) Interaction matrix and two possible assembly patterns for globally nondeterministic tiling with one species of triangles. (B) Interaction matrix and assemblies for globally deterministic but locally nondeterministic tiling with two species of triangles. (C) Assembly pathways of the (B) system starting from B triangles. Although the final structure is the same, locally nondeterminstic assemblies can have multiple pathways to reach the final structure.}
 \label{Sfig:nondet}
\end{figure*}

We clarify the definition of deterministic and nondeterministic tilings. Nondeterministic tilings are simply systems that can have more than one final assembly outcome, as in Fig.~\ref{Sfig:nondet}A. Since side 1 of the triangle can bind to either side 1 or 2 of another triangle, the system can either assemble into an `S' like structure or a hexamer. The nondeterministic nature of the system can also be predicted simply by observing the interaction matrix; the fact that side 1 can interact with two other sides suggests that depending on the kinetic pathways of assembly, there can be multiple structural outcomes. 

However, there are interaction matrices with such nondeterministic features, but with only one assembly outcome, such as the one in Fig.~\ref{Sfig:nondet}B. For this system, we find that there are multiple assembly pathways as we expect, though they all converge onto the same final assembly structure (Fig.~\ref{Sfig:nondet}C). We call such structures globally deterministic tilings since the interaction matrix encodes for a single ground state structure. However, because the assembly pathways are variable, the system is locally nondeterministic: either a triangle can bind in more than one orientation or more than one species of triangles can bind to an existing cluster, as illustrated in the second state of Fig.~\ref{Sfig:nondet}C. Therefore, we end up with three unique classes of assemblies, namely, nondeterministic, globally deterministic, and fully deterministic systems. In the main text, we use `deterministic assembly' to refer to fully deterministic systems, but in other literature, `deterministic assembly' usually refers to globally deterministic systems~\cite{Tesoro2016Apr}.

\clearpage


\section{Enumerating exhaustive list of tilings using symmetry-based method} \label{sec:tilingList}

\begin{figure}[bh!]
 \centering
 \includegraphics[width=0.75\linewidth]{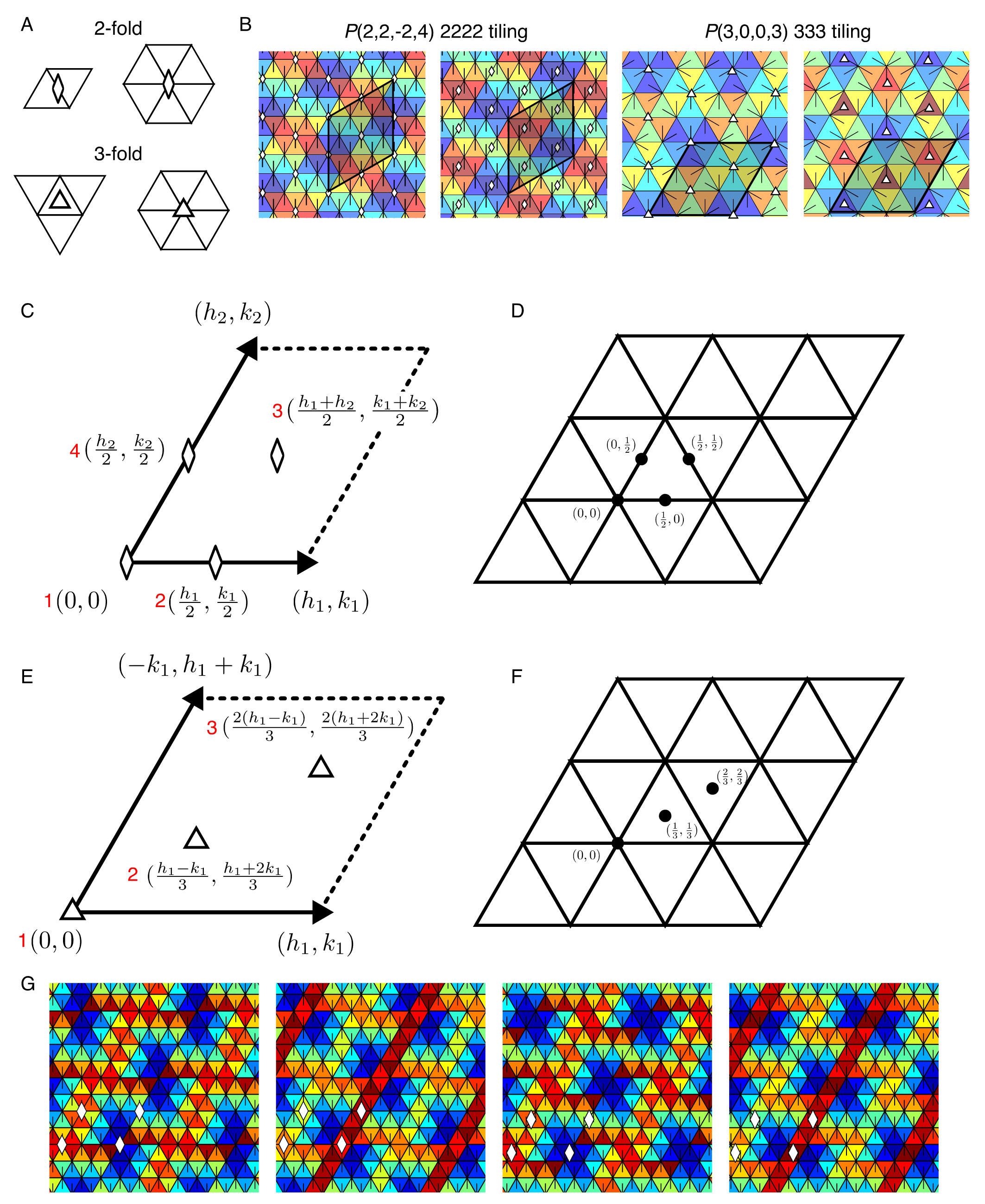}
 \caption{\textbf{Constructing variants for 2222 and 333 tilings.} (A) Possible locations for 2- and 3-fold rotational symmetry points on a triangular lattice. (B) Examples of variant tilings for 2222 and 333 tilings. (C) Locations of 2-fold rotational points on a PU cell $P(h_1,k_1,h_2,k_2)$. (D) Possible locations for the origin of the PU cell on the lattice for 2222 tilings. (E) Locations of 3-fold rotational points on a PU cell $P(h_1,k_1,h_2,k_2)$. (F) Possible locations for the origin of the PU cell on the lattice for 333 tilings. (G) An extreme example for 2222 tilings that have 4 unique variant tilings for a single PU cell.
}
 \label{sfig:variants}
\end{figure}

An important detail for the enumeration of tilings that is not described in depth in the main text is the presence of variants for 2222 and 333 tilings. Unlike 6-fold rotational symmetry points which can only lie at the vertex of the triangular lattice, 2-fold rotational symmetry points can be located either at the vertex or the middle of an edge, while 3-fold rotational symmetry points can be located either at the vertex or the middle of a triangular face (Fig.~\ref{sfig:variants}A). The freedom in choosing the location of 2- and 3-fold rotational symmetry points introduces additional tilings for some PU cells, which we refer to as variants. For example, there are two distinct 2222 tilings with PU cell $P(2,2,-2,4)$ and two distinct 333 tilings with PU cell $P(3,0,0,3)$ (Fig.~\ref{sfig:variants}B).

To identify the presence of variants, we track the locations of the rotational symmetry points on the triangular lattice. A tiling $P(h_1,k_1,h_2,k_2)$ with 2222 symmetry has 2-fold rotational symmetry points located at $(0,0)$, $\left( \frac{h_1}{2},\frac{k_1}{2}\right)$, $\left( \frac{h_1+h_2}{2},\frac{k_1+k_2}{2}\right)$, and $\left( \frac{h_2}{2},\frac{k_2}{2}\right)$ (Fig.~\ref{sfig:variants}C), where $(0,0)$ is usually considered to be on a vertex. Then, depending on the parity of $h_1$, $k_1$, $h_2$, and $k_2$, the rest of the 2-fold points either lie on a vertex or an edge, as can be seen from the value of coordinates. For example, if all PU cell components are even numbers, all symmetry points are located at the vertex (Fig.~\ref{sfig:variants}B, left). However, we sometimes obtain different tilings by moving the location of the first 2-fold point from the vertex $(0,0)$ to the middle of an edge, $(\frac{1}{2},0)$, $(0,\frac{1}{2})$, or $(\frac{1}{2},\frac{1}{2})$ (Fig.~\ref{sfig:variants}D). Referring to the example in Fig.~\ref{sfig:variants}B, shifting the first symmetry point to $(0,\frac{1}{2})$ moves all 2-fold rotational points to the edge, resulting in a distinct tiling. Similarly, the first symmetry point can be shifted to $(\frac{1}{2},0)$ or $(\frac{1}{2},\frac{1}{2})$, but in this case, the tilings turn out to be identical to already generated ones; upon proper rotation and reflection, the locations of the 2-fold points and the PU vectors can be matched to the previous tilings. 

We apply similar procedures to generate 333 tilings. The locations of the three distinct 3-fold symmetry points are $(0,0)$, $\left( \frac{h_1+h_2}{3},\frac{k_1+k_2}{3}\right)$, and $\left( \frac{2(h_1+h_2)}{3},\frac{2(k_1+k_2)}{3}\right)$ (Fig.~\ref{sfig:variants}E). For 333, the modulo of $h_1-k_1$ divided by 3 determines the location of the 3-fold points; if it is 0, both the second and third coordinate lie on the vertex, whereas if it is 1 or 2, these points lie at the center of a triangle. Similar to 2222 tilings, we generate variant tilings by shifting the 3-fold points at $(0,0)$ to $(\frac{1}{3},\frac{1}{3})$ or $(\frac{2}{3},\frac{2}{3})$ (Fig.~\ref{sfig:variants}F).

\begin{figure}[h]
 \centering
 \includegraphics[width=0.5\linewidth]{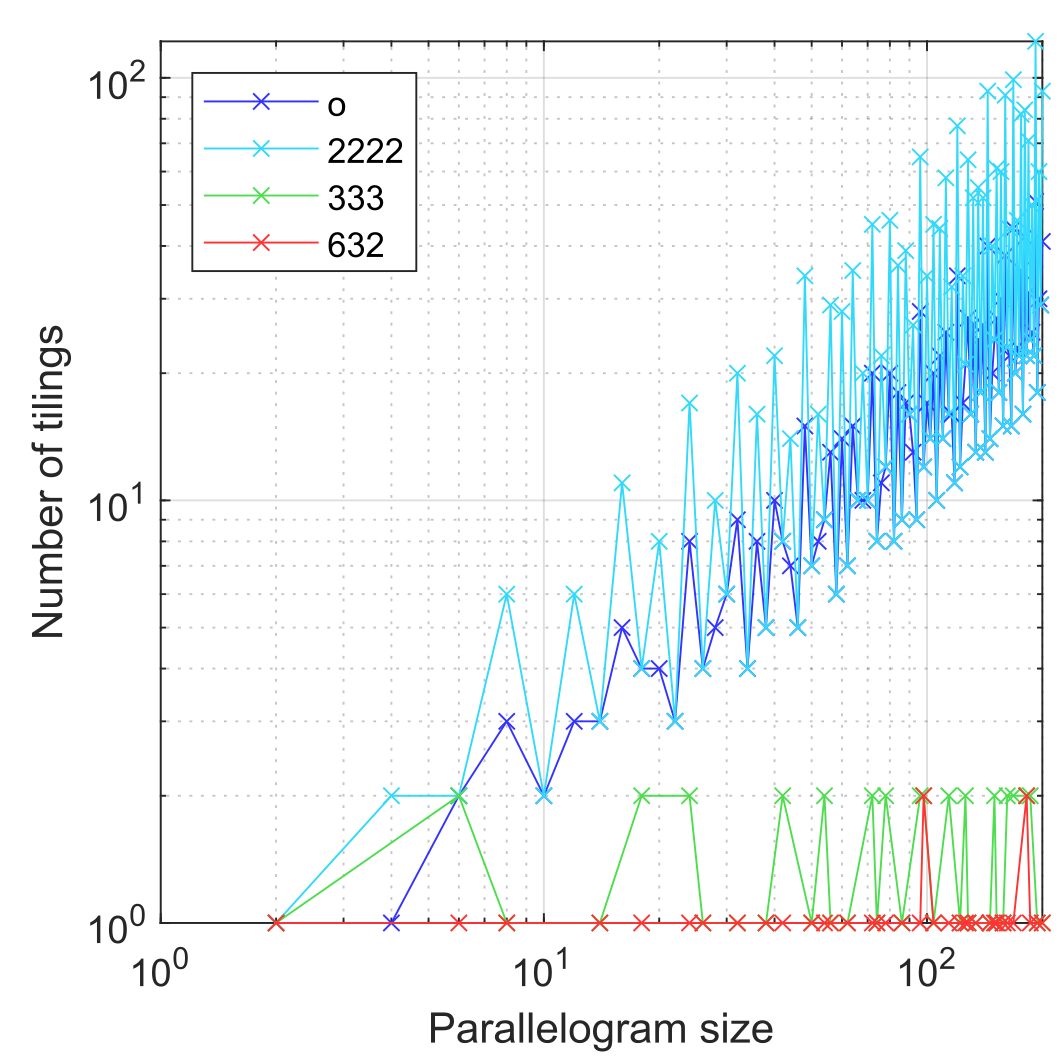}
 \caption{\textbf{Deterministic 2D planar tilings with a large number of triangle types.}
 (A) Number of tilings that are generated for a given PU cell size.
}
 \label{sfig:numPattern}
\end{figure}

Combining the coloring method explained in the main text and the variant rules explained here, we can easily generate any deterministic 2D tiling. To show the capabilities of this method, we generate an exhaustive list of 2D tilings up to $S=200$ triangles. We find 1628 o tilings, 2826 2222 tilings, 52 333 tilings, and 38 632 tilings, totaling 4544 2D tilings (Fig.~\ref{sfig:numPattern}). Examples of tilings up to 10 species of triangles for each symmetry are shown in Fig.~\ref{Sfig:tileList}. The stark difference in the number of tilings between symmetries arises from the allowed Bravais lattice for each tiling; while o and 2222 tilings can be constructed from PU cell of any two linearly independent vectors, 333 and 632 tilings can only be constructed from rhombic PU cells with opening angles of 60 degrees, or $P(h,k,-k,h+k)$. The number of parallelograms is the same between o and 2222 tilings, but we encounter more 2222 tilings due to allowed variations in the locations of 2-fold rotational symmetries. In the extreme case, a single PU cell for 2222 tilings can have four unique variant tilings (Fig.~\ref{sfig:variants}G). Similarly, there are more 333 tilings than 632 tilings owing to the variations in the allowed locations of 3-fold rotational points.

\clearpage


\section{Extension of inverse design to non-triangular tilings} \label{sec:nonTriTile}

\begin{figure}[h]
 \centering
 \includegraphics[width=0.85\linewidth]{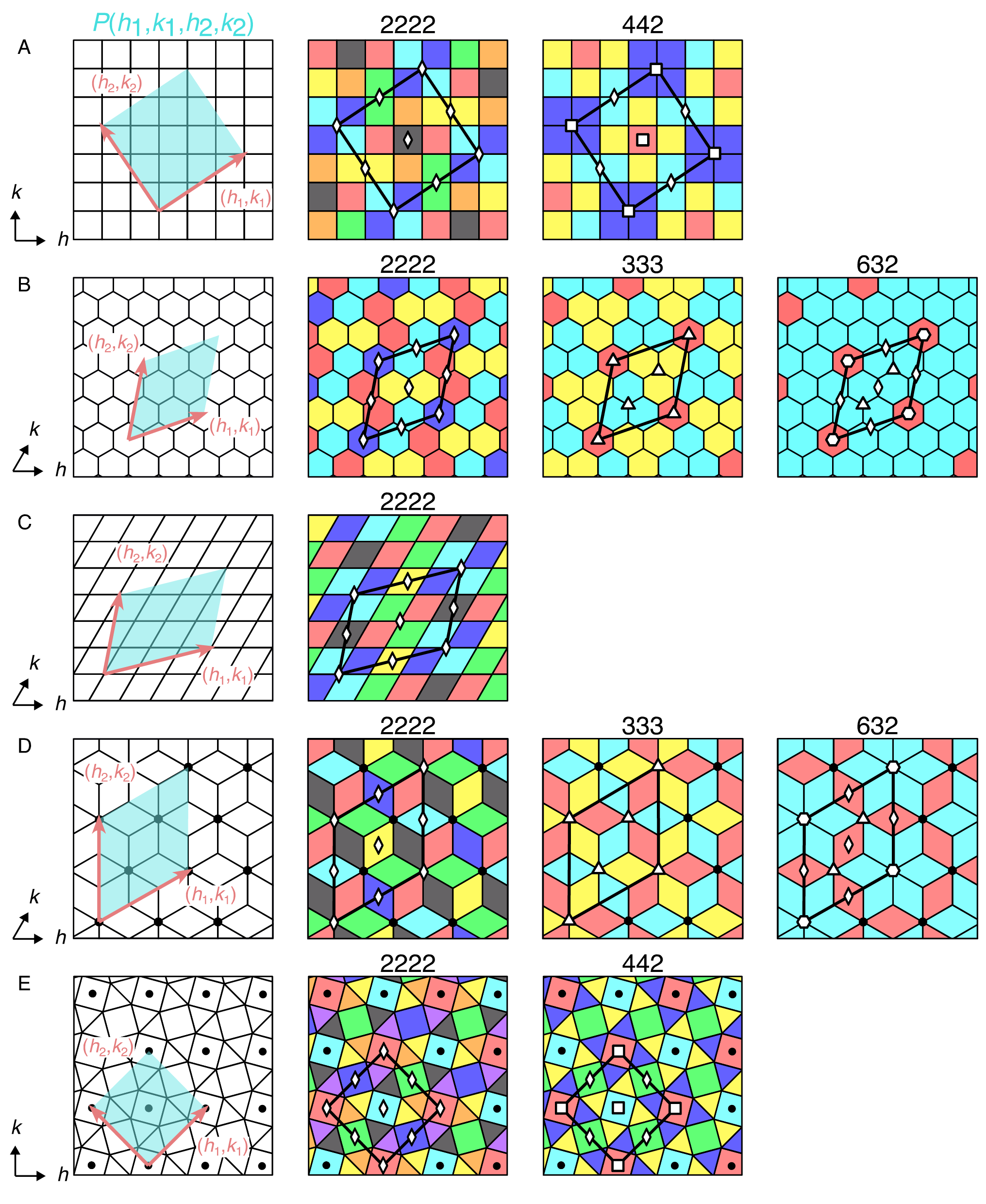}
 \caption{\textbf{Inverse design of non-triangular tiling patterns.}
Example inverse design using (A) square, (B) hexagonal, (C) parallelogram, (D) rhombille, and (D) snub square tiling. For rhombille and snub square tiling, the black dots indicate the periodicity of the blank tiling. o tilings are also possible for all examples. Orientations of the particles are also omitted but can be defined following the symmetry. 
}
 \label{sfig:nonTriTile}
\end{figure}

The inverse design techniques developed in this paper can be applied to any tilings or crystals with translational symmetry. Similar to triangular tilings, we follow the procedures below to construct deterministic, multispecies tilings. As an example, we show inverse design using square, hexagonal, parallelogram, rhombille, and snub square tiling in Fig.~\ref{sfig:nonTriTile}.

\begin{enumerate}
    \item \textbf{Find the minimal periodicity of the blank tiling.} We identify the minimal periodicity of the uncolored tiling, which can be defined using a parallelogram. We define the two linearly independent vectors that compose this parallelogram as $\textit{\textbf{h}}$ and $\textit{\textbf{k}}$. They serve as unit vectors to construct the PU cells.
    \item \textbf{Identify the PU cell with the target periodicity.} Using the coordinate system developed in 1, we identify the PU cell $P(h_1,k_1,h_2,k_2)$ that has the target periodicity. $h_1$, $k_1$, $h_2$ ,and $k_2$ all have to be integers. 
    \item \textbf{Identify the symmetries allowed by the tiling, the chosen PU cell, and the building blocks.} Possible Wallpaper groups are determined by multiple factors. Most importantly, it is dictated by ways in which the particles can bind. Since we do not allow particles to flip, we obtain Wallpaper groups with only rotational symmetry, namely o, 2222, 333, 442, and 632. Next, the tilings are also constrained by the shape of the PU cell. 432 tilings require a square PU cell, 333 and 632 a rhombic PU cell, while o and 2222 tilings can be constructed from any parallelogram shape.
    \item \textbf{Choose a symmetry group to impose on the tiling, which satisfies all the constraints given in 3.} To obtain the most economical design, choose a symmetry with the highest order of symmetry.
    \item \textbf{Identify the locations of symmetry points for a given PU cell.}  We check the compatibility with the actual tiling; every symmetry point on the tiling must be located properly such that the symmetry operation can be imposed. As described in Supplementary Section \ref{sec:tilingList}, this prohibits the placement of symmetry points at specific locations, such as a 3-fold rotational symmetry point at the middle of an edge. On the other hand, variability in the positioning of symmetry points opens up the possibility for variants, or tilings with the same PU cell but different locations of symmetry points. To search for variants, symmetry points should be shifted across every unique vertex, the center of a face, and the center of an edge.
    \item \textbf{Determine the species and orientations of particles following the symmetry.} We color in the particles and determine the orientations as described in the main text.
    \item \textbf{Derive the interaction matrix.} We determine which particle interactions to encode based on the colors and relative orientations of the adjacent particles.
\end{enumerate}

\clearpage


\section{Economy of crisscross structures}\label{sec:criss}

\begin{figure}[h]
 \centering
 \includegraphics[width=0.8\linewidth]{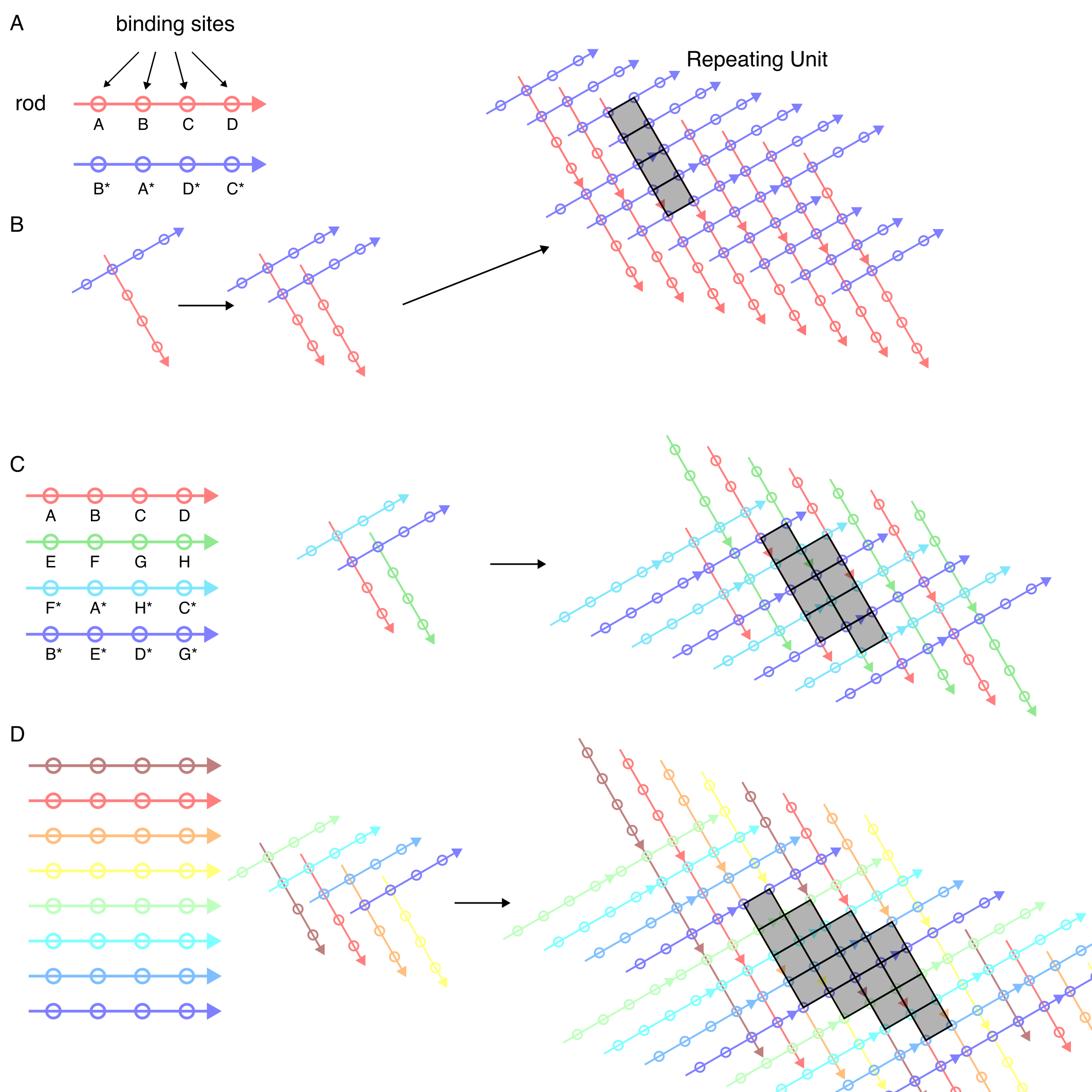}
 \caption{\textbf{An example design of crisscross structures with valence four slats.}
 (A) Design of two species slats of valence four. Species are illustrated by unique colors, whereas the binding site along the slat is illustrated as a circle. Complementary interactions are denoted using an asterisk. (B) The assembly structure of the two species crisscross structure. The repeating unit is highlighted in gray. Extensions to four (C) and eight (D) species slats are illustrated. 
}
 \label{sfig:slatDesign}
\end{figure}

In contrast to the conventional tiling schemes discussed in this paper, the crisscross crystals are assembled using slats with multiple binding sites along their length~\cite{Minev2021Mar, Wintersinger2023Mar}. As an example, we consider two species of slats, each with four binding sites, which we refer to as valence, $V$ (Fig.~\ref{sfig:slatDesign}A). For each binding site, we assign specific interactions which are denoted by capital letters. The interactions are assigned such that all bonds are satisfied when one slat binds to four other slats of different species. As a result, the slats assemble in a slightly staggered manner, and the final 2D crystal consists of two orthogonal layers of slats, in which each layer consists of the same species of slats (Fig.~\ref{sfig:slatDesign}B).

Similar to the tiling system, increasing the number of unique slats leads to crystals with larger periodicity. We characterize the area of the repeating unit, $A$, in the 2D crystal using a representative length scale set by the length of the slat, $L$. In the two species case, we obtain $A=\frac{L^2}{4}$, shown in Fig.~\ref{sfig:slatDesign}B. Further, the crisscross crystal can be made arbitrarily complex by increasing the number of unique slats in the system (Fig.~\ref{sfig:slatDesign}C and D). In short, as the number of unique slats increases, the periodicity of the slats in both the top and the bottom layer of the crisscross increases, leading to larger periodicity. For the examples shown, the area of the repeating unit is $A=\frac{L^2}{2}$ and $A=L^2$ for four and eight species of slats, respectively.

\begin{figure}[h]
 \centering
 \includegraphics[width=0.9\linewidth]{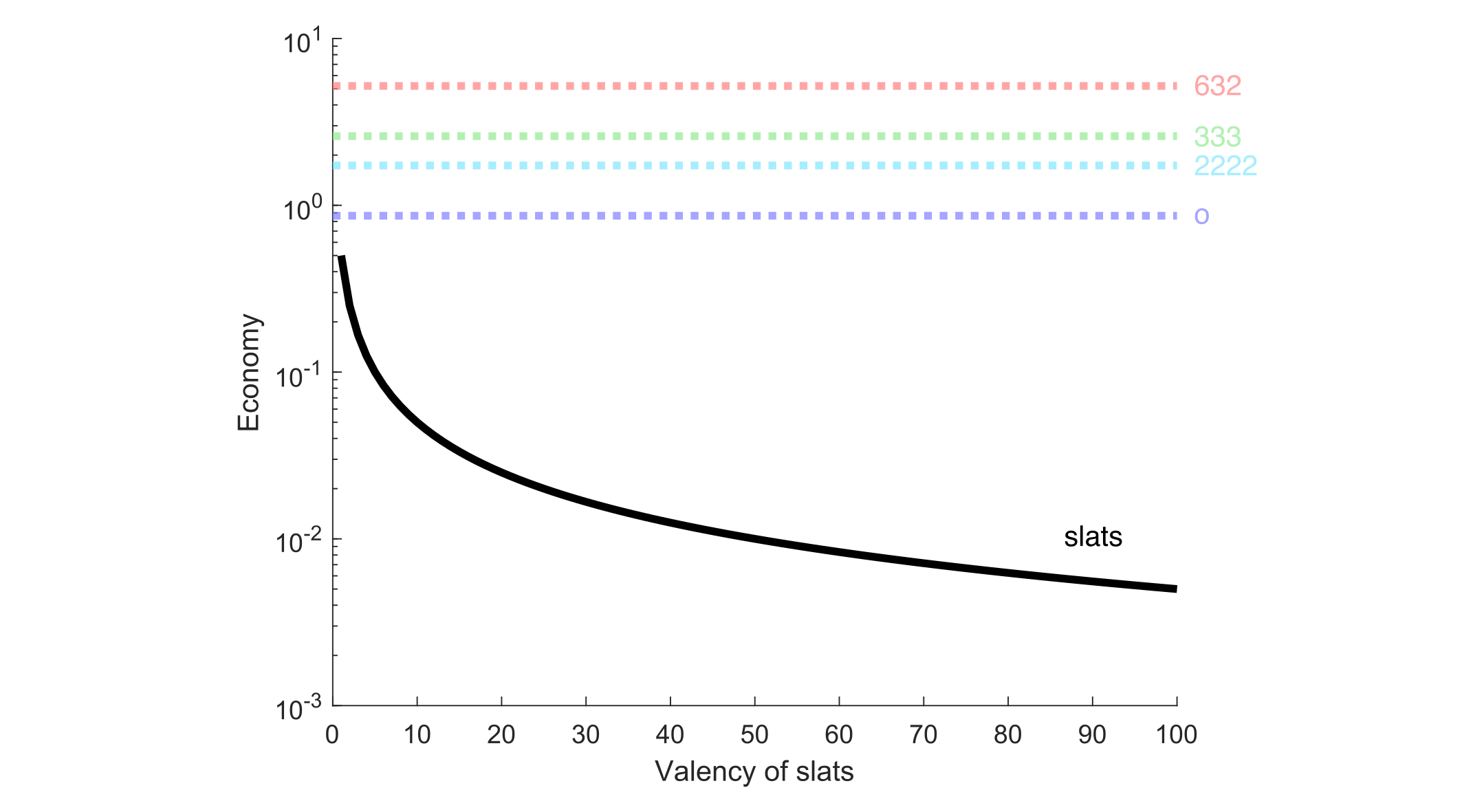}
 \caption{\textbf{Economy of crisscross structures compared against that of triangular tilings.}
}
 \label{sfig:slats}
\end{figure}

So how does the economy of the crisscross structure compare with the conventional tilings? To answer this question, we redefine the economy to allow comparison between the two systems. The basic concept is similar to what is described in the main text; we define the `cost' of assembly as the number of unique components, $N$, and the `value' as the area of the repeating unit, $A$. Here, we distinguish $A$, which has units of length squared, from the PU cell size, $S$, which has units of subunits, described in the main text. Using the slat length, $L$, the repeating  crisscross structure in Fig.~\ref{sfig:slatDesign} satisfies the equation
\begin{equation}
    A=\frac{NL^2}{8}.
\end{equation}
In general, for a crisscross system with slats of valence $V$, the area of the repeating unit is
\begin{equation}
    A=\frac{NL^2}{2V}.
\end{equation}
Defining the economy as value over cost normalized by the representative length squared, we obtain
\begin{equation}
    E_\textrm{slat}=\frac{A}{NL^2}=\frac{1}{2V}.
\end{equation}
In contrast, using the asymptotic equation for triangle tilings discussed in the main text $S=NO$, the economy of triangle tilings is given by
\begin{equation}
    E_\textrm{tiling}=\frac{A}{NL^2}=\frac{\sqrt{3}}{2}O,
\end{equation}
where $L$ is the edge length of a triangle. Comparison between the economies of the two systems shows that triangle tilings are more economical than the crisscross crystals. We note that in both cases, the economy is independent of the number of unique subunits in the system. However, in the case of crisscross crystals, the economy decreases inverse proportionally as the valency of the slats increases (Fig.~\ref{sfig:slats}). We conclude by highlighting an interesting trade-off between the economy and the nucleation barrier in the crisscross systems; higher valency allows for heightening the nucleation barrier, which prohibits spurious nucleation but, in turn, reduces the system economy. 

\section{ssDNA handle design}\label{sec:handles}

\begin{figure}[h]
 \centering
 \includegraphics[width=0.6\linewidth]{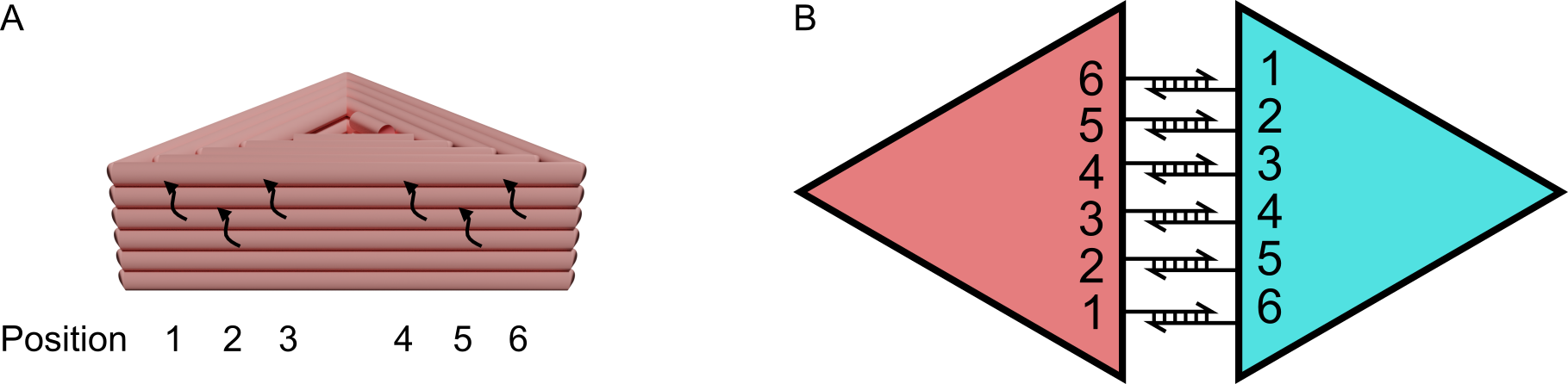}
 \caption{\textbf{Handle positions on the DNA origami triangles} (A) Position of the handles on the DNA origami triangles. (B) Strands on Position 1 bind to Position 6, Position 2 binds to Position 5, and Position 3 binds to Position 4.}
 \label{fig:handles}
\end{figure}

Following the strand design protocols established by Seeman~\cite{seeman1990sequence}, we design ssDNA strands to program interactions between edges of DNA origami triangles, as summarized in Table~\ref{tab:sidesequence}. In short, a unique set of ssDNA sequences is generated from a library of `vocabularies' composed of five nucleic acid letters. The method allows for the generation of a large set of orthogonal sequences with minimal crosstalk. Here, we generated 72 unique 6 base pair (bp) sequences and their complementary sequences, totaling 144 sequences. We combine six of these strands to encode for the interaction on each side of the triangle, whose locations are illustrated in Fig.~\ref{fig:handles}. The combinations of strands are chosen such that the sum of the binding free energies of the six strands calculated using the nearest neighbor model is around $-30$ kcal$/$mol~\cite{SantaLucia_1996Jan}. Although the total number of unique edges that can be encoded using 144 unique sequences is only 24, we generate a far larger library of edge interactions by changing the locations of the sequences. For example, interactions A and J use the same set of sequences, but their positions are altered. The combinations of sequences for each assembly experiment are shown in Table~\ref{tab:triInteractions} and the corresponding interaction matrices in Fig.~\ref{Sfig:intMatList}. Though the same sequence strands are repeated for different interactions, we do not observe any unwanted crosstalks in the assembly experiment. A more detailed design method for the sequences is described in~\cite{videbaek2023economical}.

\begin{figure}[ht]
 \centering
 \includegraphics[width=0.65\textwidth]{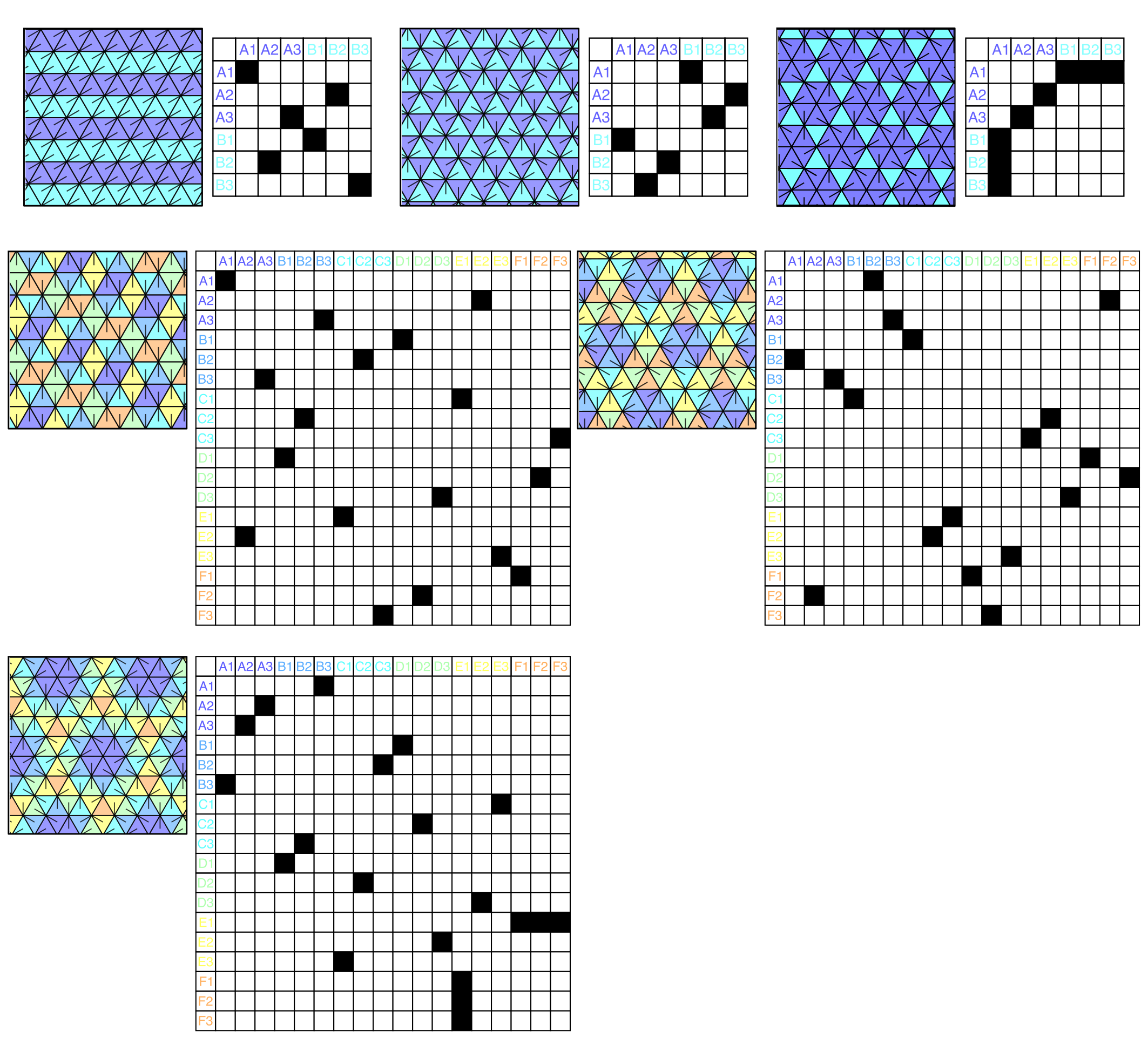}
 \caption{\textbf{Interaction matrices of tilings assembled in experiment.}}
\end{figure}
\begin{figure}[ht]
 \addtocounter{figure}{-1}
 \centering
 \includegraphics[width=0.9\textwidth]{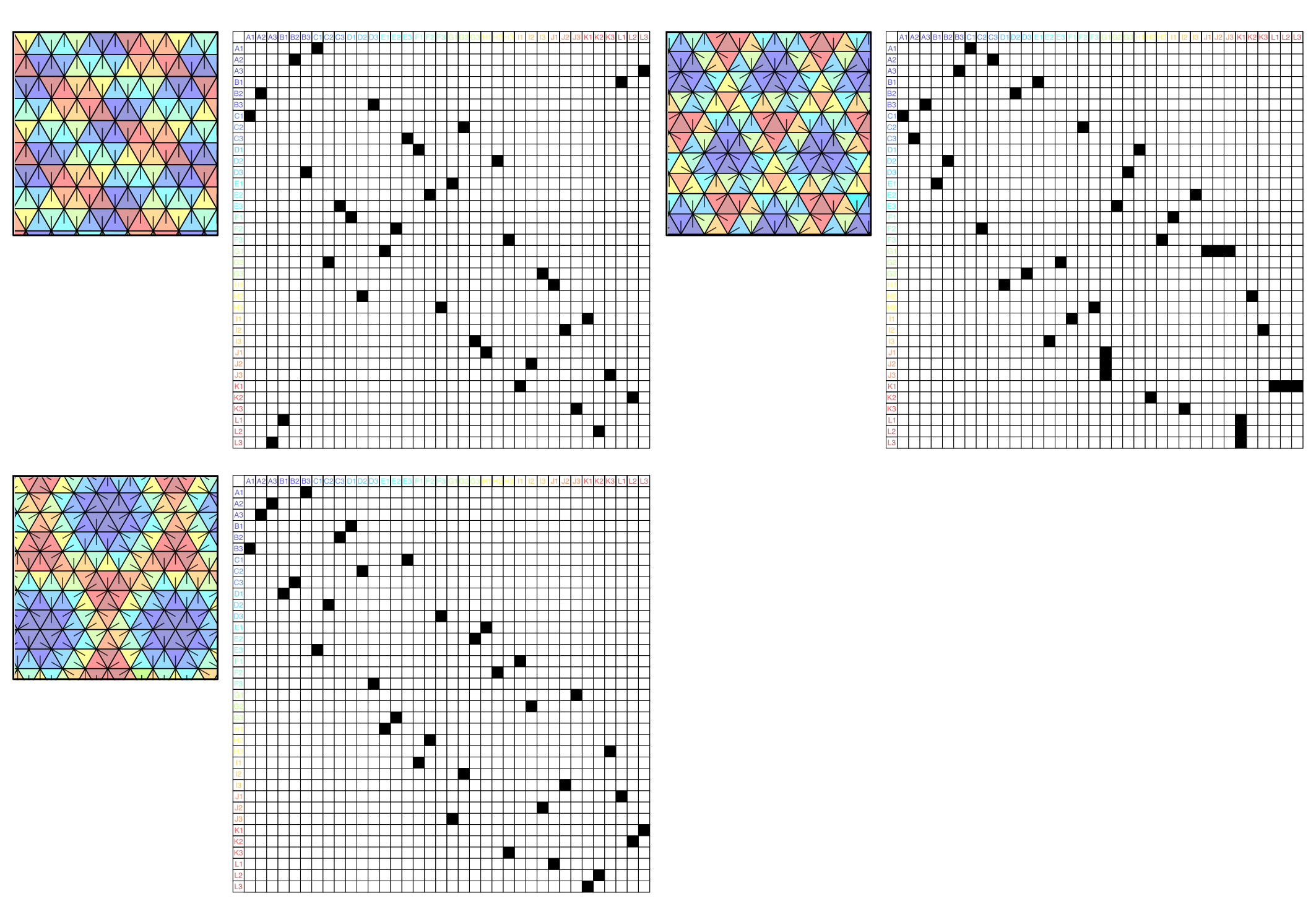}
 \caption{\textbf{Interaction matrices of tilings assembled in experiment (continued).}}
 \label{Sfig:intMatList}
\end{figure}

\clearpage


\begin{longtable}{p{0.09\linewidth} p{0.09\linewidth} p{0.09\linewidth} p{0.09\linewidth}p{0.09\linewidth} p{0.09\linewidth} p{0.09\linewidth} p{0.12\linewidth}}
\caption{\textbf{DNA sequence for multispecies assemblies.} A list of the set of six interaction sequences that make up a side interaction of a monomer and an estimate of their binding free energy. As shown in Fig.~\ref{fig:handles}, Position 1 binds to Position 6, Position 2 binds to Position 5, and Position 3 binds to Position 4. A lowercase `s' indicates that the sequence is self-complementary. An asterisk denotes a sequence that is complementary to the original.} \label{tab:sidesequence} \\
Name & Position 1 & Position 2 & Position 3 & Position 4 & Position 5 & Position 6 & $\Delta$G [kcal$/$mol] \\
\hline
\endfirsthead

Name & Position 1 & Position 2 & Position 3 & Position 4 & Position 5 & Position 6 & $\Delta$G [kcal$/$mol] \\
\hline 
\endhead

\endfoot

\endlastfoot
sA	&	ACTAGC	& 	AGTTAC	&	TAGTCT	&	AGACTA	&	GTAACT	&	GCTAGT	&	-30.57	\\
sB	&	TTAACC	& 	TCGACA	&	TTGGAT	&	ATCCAA	&	TGTCGA	&	GGTTAA	&	-30.23	\\
sC	&	TCAGAC	& 	GTCTAG	&	TACCTT	&	AAGGTA	&	CTAGAC	&	GTCTGA	&	-30.17	\\
sD	&	GATCTT	& 	CTGATC	&	TCCACA	&	TGTGGA	&	GATCAG	&	AAGATC	&	-31.27	\\

A	&	CAATAG	& 	TGATTG	&	CTAGGA	&	CACATC	&	ACGAAG	&	ACCTGA	&	-31.38	\\
B	&	ATGACA	& 	TACAGG	&	AACCTA	&	GAGACA	&	GACAGA	&	ACTAAC	&	-30.57	\\
C	&	GTACAT	& 	AGTCAG	&	CGATGG	&	CTTACT	&	AGTATC	&	GTATGT	&	-31.16	\\
D	&	GCATCT	& 	TATTCC	&	AGATTC	&	TTCTCA	&	CTGTGA	&	TCGTAC	&	-30.72	\\
E	&	AGATAG	& 	TTCCTG	&	TTCCAT	&	GATATG	&	ATGCAC	&	AACATT	&	-30.31	\\
F	&	ACAATT	& 	CGTCCA	&	CTTGTA	&	CTACAC	&	GACAGA	&	AACTAT	&	-30.58	\\
G	&	AGTTCC	& 	CGATTA	&	ATTCTG	&	ATTCAG	&	CTTGAG	&	GTAGAT	&	-30.27	\\
H	&	GGATAA	& 	TCATCC	&	GGTATT	&	GGTAAT	&	ACTGAG	&	AGAGAT	&	-29.84	\\
I	&	TTGGCA	& 	GACCTC	&	CCTATG	&	CTTAGG	&	TAACAG	&	TCTTCT	&	-31.1	\\
J	&	CACATC	& 	CAATAG	&	ACGAAG	&	TGATTG	&	ACCTGA	&	CTAGGA	&	-31.38	\\
K	&	GAGACA	& 	ATGACA	&	GACAGA	&	TACAGG	&	ACTAAC	&	AACCTA	&	-30.57	\\
L	&	CTTACT	& 	GTACAT	&	AGTATC	&	AGTCAG	&	GTATGT	&	CGATGG	&	-31.16	\\
M	&	TTCTCA	& 	GCATCT	&	CTGTGA	&	TATTCC	&	TCGTAC	&	AGATTC	&	-30.72	\\
N	&	GATATG	& 	AGATAG	&	ATGCAC	&	TTCCTG	&	AACATT	&	TTCCAT	&	-30.31	\\
O	&	CTACAC	& 	ACAATT	&	GACAGA	&	CGTCCA	&	AACTAT	&	CTTGTA	&	-30.58	\\
P	&	ATTCAG	& 	AGTTCC	&	CTTGAG	&	CGATTA	&	GTAGAT	&	ATTCTG	&	-30.27	\\
Q	&	GGTAAT	& 	GGATAA	&	ACTGAG	&	TCATCC	&	AGAGAT	&	GGTATT	&	-29.84	\\
R	&	CTTAGG	& 	TTGGCA	&	TAACAG	&	GACCTC	&	TCTTCT	&	CCTATG	&	-31.1	\\
S	&	ACGAAG	& 	CTAGGA	&	ACCTGA	&	CAATAG	&	CACATC	&	TGATTG	&	-31.38	\\
T	&	GACAGA	& 	AACCTA	&	ACTAAC	&	ATGACA	&	GAGACA	&	TACAGG	&	-30.57	\\
U	&	AGTATC	& 	CGATGG	&	GTATGT	&	GTACAT	&	CTTACT	&	AGTCAG	&	-31.16	\\
V	&	CTGTGA	&	AGATTC	&	TCGTAC	&	GCATCT	&	TTCTCA	&	TATTCC	&	-30.72	\\
W	&	ATGCAC	&	TTCCAT	&	AACATT	&	AGATAG	&	GATATG	&	TTCCTG	&	-30.31	\\
X	&	GACAGA	&	CTTGTA	&	AACTAT	&	ACAATT	&	CTACAC	&	CGTCCA	&	-30.58	\\
Y	&	CTTGAG	&	ATTCTG	&	GTAGAT	&	AGTTCC	&	ATTCAG	&	CGATTA	&	-29.84	\\

A*	&	TCAGGT	& 	CTTCGT	&	GATGTG	&	TCCTAG	&	CAATCA	&	CTATTG	&	-31.38	\\
B*	&	GTTAGT	& 	TCTGTC	&	TGTCTC	&	TAGGTT	&	CCTGTA	&	TGTCAT	&	-30.57	\\
C*	&	ACATAC	& 	GATACT	&	AGTAAG	&	CCATCG	&	CTGACT	&	ATGTAC	&	-31.16	\\
D*	&	GTACGA	& 	TCACAG	&	TGAGAA	&	GAATCT	&	GGAATA	&	AGATGC	&	-30.72	\\
E*  &	AATGTT	& 	GTGCAT	&	CATATC	&	ATGGAA	&	CAGGAA	&	CTATCT	&	-30.31	\\
F*	&	ATAGTT	& 	TCTGTC	&	GTGTAG	&	TACAAG	&	TGGACG	&	AATTGT	&	-30.58	\\
G*	&	ATCTAC	& 	CTCAAG	&	CTGAAT	&	CAGAAT	&	TAATCG	&	GGAACT	&	-30.27	\\
H*	&	ATCTCT	& 	CTCAGT	&	ATTACC	&	AATACC	&	GGATGA	&	TTATCC	&	-29.84	\\
I*	&	AGAAGA	& 	CTGTTA	&	CCTAAG	&	CATAGG	&	GAGGTC	&	TGCCAA	&	-31.1	\\
J*	&	TCCTAG	& 	TCAGGT	&	CAATCA	&	CTTCGT	&	CTATTG	&	GATGTG	&	-31.38	\\
K*	&	TAGGTT	& 	GTTAGT	&	CCTGTA	&	TCTGTC	&	TGTCAT	&	TGTCTC	&	-30.57	\\
L*	&	CCATCG	& 	ACATAC	&	CTGACT	&	GATACT	&	ATGTAC	&	AGTAAG	&	-31.16	\\
M*	&	GAATCT	& 	GTACGA	&	GGAATA	&	TCACAG	&	AGATGC	&	TGAGAA	&	-30.72	\\
N*	&	ATGGAA	& 	AATGTT	&	CAGGAA	&	GTGCAT	&	CTATCT	&	CATATC	&	-30.31	\\
O*	&	TACAAG	& 	ATAGTT	&	TGGACG	&	TCTGTC	&	AATTGT	&	GTGTAG	&	-30.58	\\
P*	&	CAGAAT	& 	ATCTAC	&	TAATCG	&	CTCAAG	&	GGAACT	&	CTGAAT	&	-30.27	\\
Q*	&	AATACC	& 	ATCTCT	&	GGATGA	&	CTCAGT	&	TTATCC	&	ATTACC	&	-29.84	\\
R*	&	CATAGG	& 	AGAAGA	&	GAGGTC	&	CTGTTA	&	TGCCAA	&	CCTAAG	&	-31.1	\\
S*	&	CAATCA	& 	GATGTG	&	CTATTG	&	TCAGGT	&	TCCTAG	&	CTTCGT	&	-31.38	\\
T*	&	CCTGTA	& 	TGTCTC	&	TGTCAT	&	GTTAGT	&	TAGGTT	&	TCTGTC	&	-30.57	\\
U*	&	CTGACT	& 	AGTAAG	&	ATGTAC	&	ACATAC	&	CCATCG	&	GATACT	&	-31.16	\\
V*	&	GGAATA	&	TGAGAA	&	AGATGC	&	GTACGA	&	GAATCT	&	TCACAG	&	-30.72	\\
W*	&	CAGGAA	&	CATATC	&	CTATCT	&	AATGTT	&	ATGGAA	&	GTGCAT	&	-30.31	\\
X*	&	TGGACG	&	GTGTAG	&	AATTGT	&	ATAGTT	&	TACAAG	&	TCTGTC	&	-30.58	\\
Y*	&	TAATCG	&	CTGAAT	&	GGAACT	&	ATCTAC	&	CAGAAT	&	CTCAAG	&	-29.84	\\
\end{longtable}

\begin{longtable}{p{0.09\linewidth} p{0.18\linewidth} p{0.10\linewidth} p{0.15\linewidth}p{0.15\linewidth} p{0.15\linewidth}}
\caption{\textbf{Triangle interactions.} This table enumerates the different side interactions from Table~\ref{tab:sidesequence} used to generate the tilings for the assembly experiment. `L' indicates triangles that were labeled with gold nanoparticles.} \label{tab:triInteractions} \\
Symmetry &  Number of subunits	&	Subunit ID	&	Side 1 strands	&	Side 2 strands	&	Side 3 strands	\\
\hline
\endfirsthead

Symmetry &  Number of subunits	&	Subunit ID	&	Side 1 strands	&	Side 2 strands	&	Side 3 strands	\\
\hline
\endhead

\endfoot
\endlastfoot
632 &   N=1	&	1	&	sA	&	sA	&	sA  \\
2222    &  N=2	&	1(L)	&	sA	&	B	&	sB    \\
        &  	    &	2	&	sC	&	B*	&	sD    \\
2222    &  N=6	&	1(L)	&	sA	&	B	&	C    \\
        &  	    &	2	&	D	&	E	&	C*    \\
        &  	    &	3	&	G	&	E*	&	I    \\
        &  	    &	4	&	D*	&	K	&	sB    \\
        &  	    &	5	&	G*	&	B*	&	sD    \\
        &  	    &	6	&	sC	&	K*	&	I*    \\
2222    &  N=12	&	1(L)	&	A	&	B	&	C    \\
        &  	    &	2	&	D	&	B*	&	F    \\
        &  	    &	3	&	A*	&	E	&	I    \\
        &  	    &	4	&	G	&	H	&	F*    \\
        &  	    &	5	&	J	&	K	&	I*    \\
        &  	    &	6	&	G*	&	K*	&	L    \\     
        &  	    &	7	&	J*	&	E*	&	O    \\
        &  	    &	8	&	M	&	H*	&	L*    \\
        &  	    &	9	&	P	&	N	&	O*    \\
        &  	    &	10	&	M*	&	N*	&	R    \\
        &  	    &	11(L)	&	P*	&	Q	&	R*    \\
        &  	    &	12	&	D*	&	Q*	&	C*    \\
333 &  N=2	&	1(L)	&	A	&	B	&	C    \\
    &  	    &	2	&	A*	&	C*	&	B*    \\
333 &  N=6	&	1(L)	&	A	&	K	&	F    \\
    &  	    &	2	&	J	&	A*	&	F*    \\
    &  	    &	3	&	J*	&	H	&	C    \\
    &  	    &	4	&	D	&	N	&	I    \\
    &  	    &	5	&	C*	&	H*	&	I*    \\
    &  	    &	6	&	D*	&	K*	&	N*    \\
333 &  N=12	&	1(L)	&	P	&	H	&	R    \\
    &  	    &	2	&	S	&	Q	&	R*    \\
    &  	    &	3	&	P*	&	T	&	H*    \\
    &  	    &	4	&	V	&	Q*	&	U    \\
    &  	    &	5	&	S*	&	K	&	N*    \\
    &  	    &	6	&	Y	&	T*	&	X    \\
    &  	    &	7	&	A	&	N	&	U*    \\
    &  	    &	8(L)	&	V*	&	W	&	X*    \\
    &  	    &	9	&	Y*	&	E	&	K*    \\
    &  	    &	10	&	A*	&	A*	&	A*    \\
    &  	    &	11	&	D	&	W*	&	E*    \\
    &  	    &	12	&	D*	&	D*	&	D*    \\
632 &  N=2	&	1(L)	&	A	&	E	&	E*    \\
    &  	    &	2	&	A*	&	A*	&	A*    \\
632 &  N=6	&	1(L)	&	D	&	C*	&	C    \\
    &  	    &	2	&	J	&	E	&	D*    \\
    &  	    &	3	&	I*	&	K	&	E*    \\
    &  	    &	4	&	J*	&	K*	&	H*    \\
    &  	    &	5	&	A	&	H	&	I    \\
    &  	    &	6	&	A*	&	A*	&	A*    \\
632 &  N=12	&	1(L)	&	D	&	C*	&	C    \\
    &  	    &	2	&	J	&	E	&	D*    \\
    &  	    &	3	&	I*	&	K	&	E*    \\
    &  	    &	4	&	J*	&	K*	&	R    \\
    &  	    &	5	&	M	&	N	&	I    \\
    &  	    &	6	&	P	&	Q	&	R*    \\
    &  	    &	7	&	L*	&	T	&	N*    \\
    &  	    &	8(L)	&	M*	&	Q*	&	U    \\
    &  	    &	9	&	P*	&	T*	&	H*    \\
    &  	    &	10	&	S	&	H	&	L    \\
    &  	    &	11	&	O*	&	W	&	U*    \\
    &  	    &	12	&	S*	&	W*	&	O    \\

\end{longtable}

\clearpage


\section{Fourier transform of TEM images}\label{sec:fourier}

\begin{figure}[ht]
 \centering
 \includegraphics[width=0.99\textwidth]{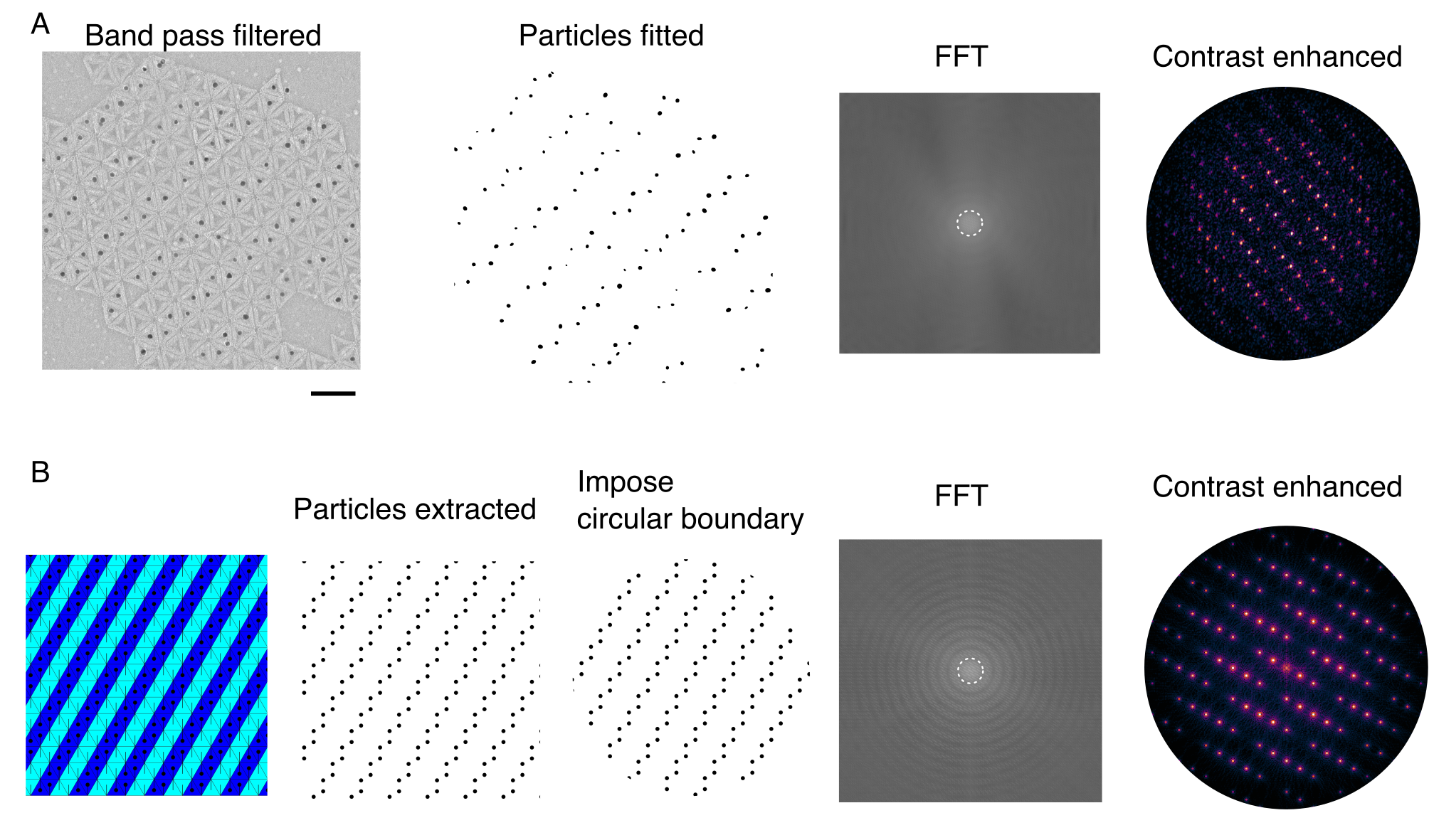}
 \caption{\textbf{Image processing procedures for Fourier transform of the gold nanoparticle pattern.} Image processing procedures for gold nanoparticle patterns for (A) TEM micrographs and (B) ideal lattice. The dotted circle region in the first FFT image is cropped and contrast-enhanced to obtain the final FFT pattern.}
 \label{Sfig:imageprocess}
\end{figure}

We characterize the patterns of gold nanoparticles on the triangular lattices using Fourier transforms. To analyze the TEM micrographs, we bandpass filter the micrograph and threshold to isolate the gold nanoparticle on the lattice. The contrast difference between the background and the gold nanoparticle allows fairly simple thresholding after bandpass filtering. We further fit the gold nanoparticle with ellipses using Fiji~\cite{Schindelin2012Jul}. Due to the thresholding procedure, some gold nanoparticles seem to change in size and circularity. However, local changes in the gold nanoparticle geometry do not impact the low-spatial-frequency Fourier transform patterns. Finally, we take a fast Fourier transform (FFT) of the fitted ellipses using Fiji and enhance the contrast near the center. To focus on the low-wavelength peaks, the transformed images are cropped near the center, along the first minimum of the enveloping Bessel function that corresponds to the FFT of the gold nanoparticles. The flow for image analysis is summarized in Fig.~\ref{Sfig:imageprocess}A.

For comparison, we simulate the coordinates of an ideal lattice and the positions of gold nanoparticles. Shown Fig.~\ref{Sfig:imageprocess}B, we simulate the locations of each species of the triangles on a lattice, assuming perfectly fitting equilateral triangles, 50~nm in size. The center of the triangles with target species is labeled with gold nanoparticles, corresponding to 10~nm in size. Similar to the TEM micrograph analysis, we extract the positions of the gold nanoparticles to Fourier transform. To reduce the boundary effect in the FFT, we impose circular masks on the gold nanoparticle patterns. Finally, we obtain the Fourier transform patterns and the center of the pattern is enhanced in contrast. The FFT patterns are rotated accordingly to match the orientation for comparison.

\begin{figure}[!t]
 \centering
 \includegraphics[width=0.8\textwidth]{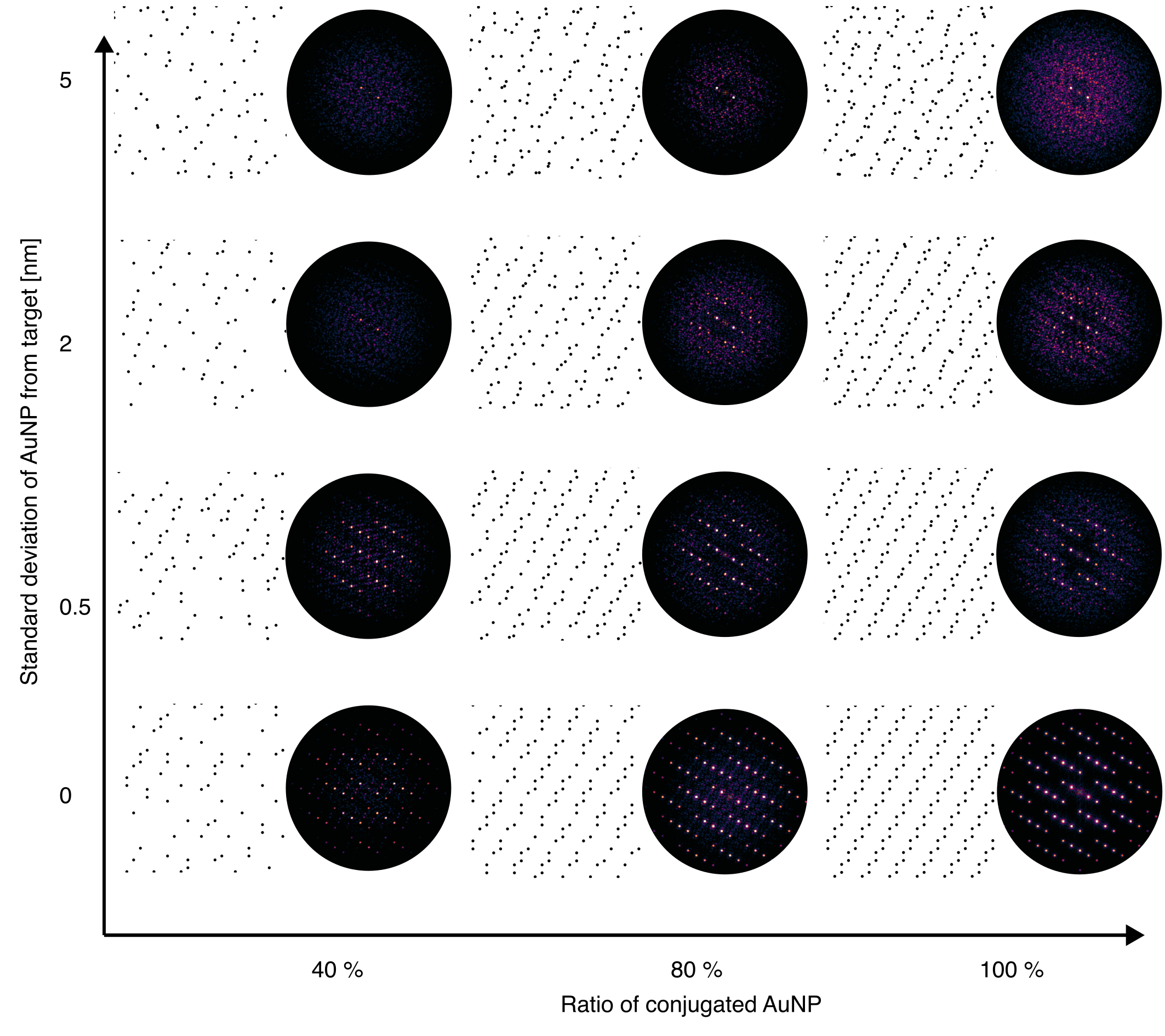}
 \caption{\textbf{The gold nanoparticle patterns and respective FFT patterns for reduced conjugation and localization.}}
 \label{Sfig:fftblur}
\end{figure}

Unique peak patterns of the FFT near the center are visible for a relatively wide range of gold nanoparticle conjugation ratios, as long as the fluctuations of the gold nanoparticles are below 2~nm. For the 2 species tiling in Fig.~\ref{Sfig:imageprocess}B, we generate gold nanoparticle patterns with conjugation ratios varying from 40\% to 100\% and the standard deviation of fluctuations from 0~nm to 5~nm (Fig.~\ref{Sfig:fftblur}). We find that both reducing the conjugation and increasing the fluctuation blur the FFT peaks starting far from the center. The central peak patterns can still be discerned for conjugation ratio above 40\% and standard deviation of fluctuations below 2~nm. In the experiment, we typically find that the conjugation ratio is above 75\%. We suspect that the fluctuation of gold nanoparticles from the center of the triangle in the experiment is smaller than 2~nm. Other additional factors not considered here include free gold nanoparticles not conjugated on the lattice and systematic deformation of the triangular lattice.

\begin{figure}[ht]
 \centering
 \includegraphics[width=0.99\textwidth]{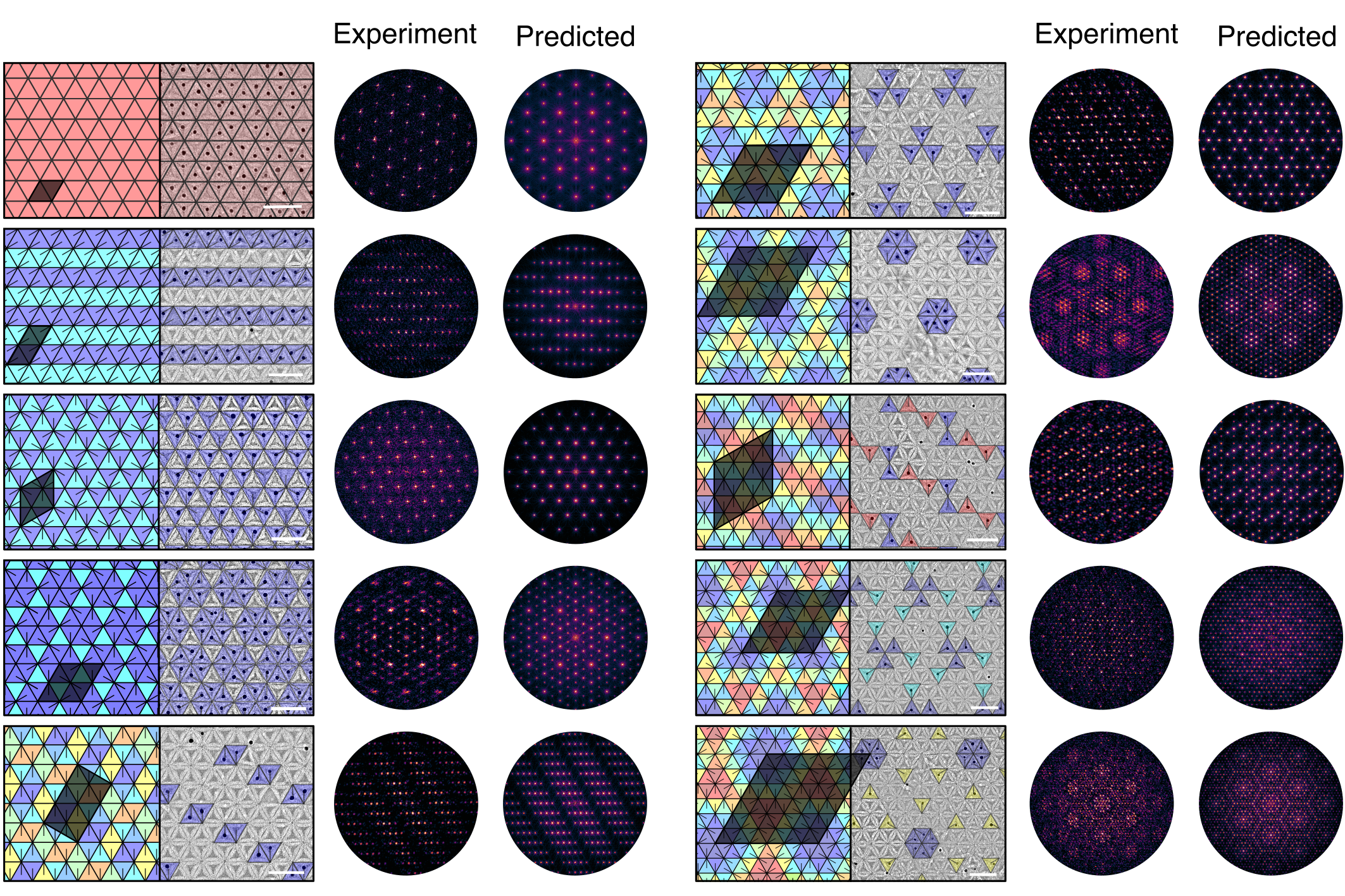}
 \caption{\textbf{Summary of all assembly experiments and the corresponding FFT patterns.} The prescribed tiling patterns and the corresponding experimental results are shown. For each tiling, the FFT patterns of the gold nanoparticles are shown, along with the predicted FFT patterns. All scale bars are 100 nm.}
 \label{Sfig:fftExp}
\end{figure}
\clearpage

\bibliography{main.bib}
\clearpage


\begin{figure}[ht]
 \centering
 \includegraphics[width=0.9\textwidth]{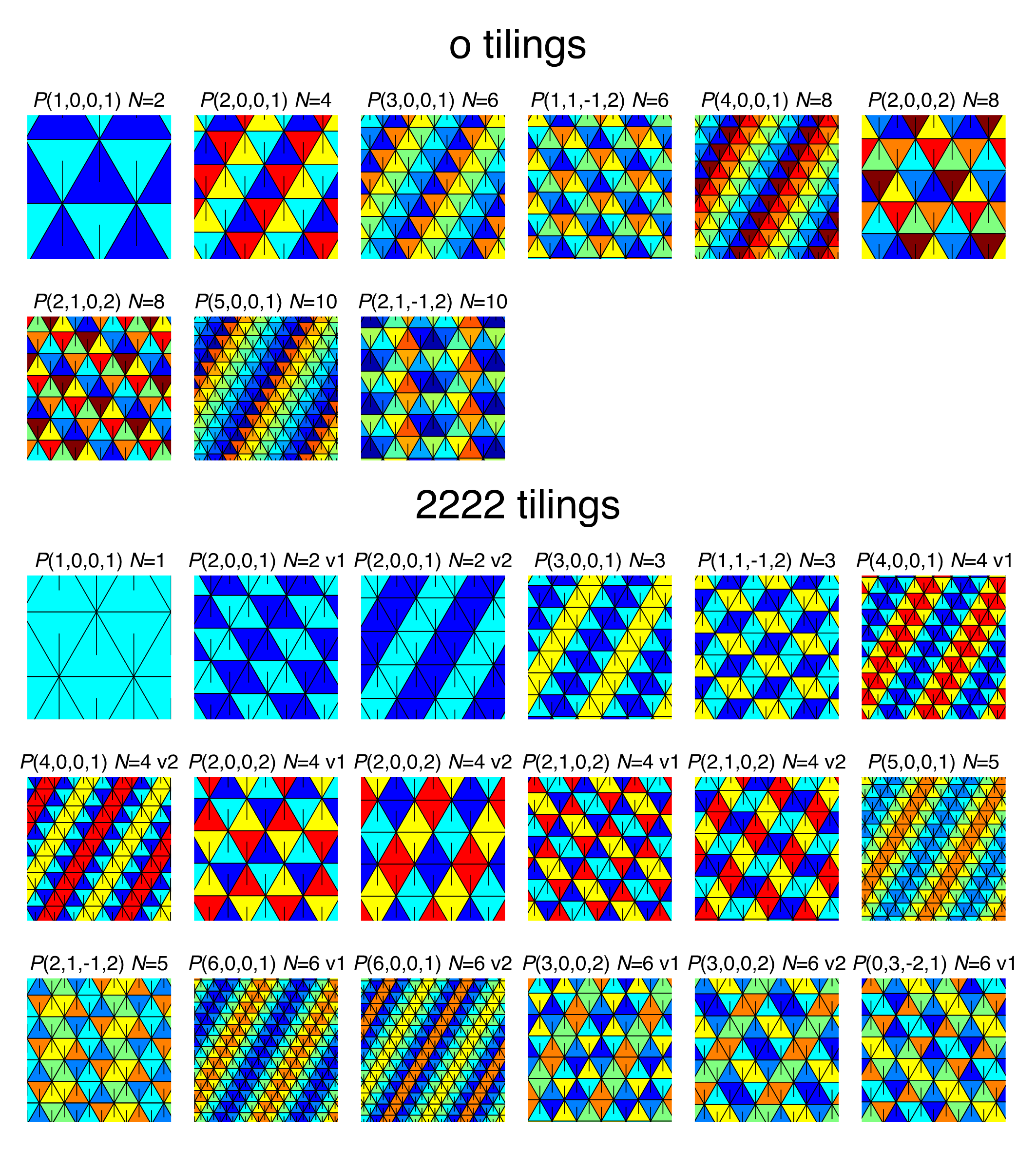}
 \caption{\textbf{List of tilings up to 10 species of triangles.}}
\end{figure}
\begin{figure}[ht]
 \addtocounter{figure}{-1}
 \centering
 \includegraphics[width=0.9\textwidth]{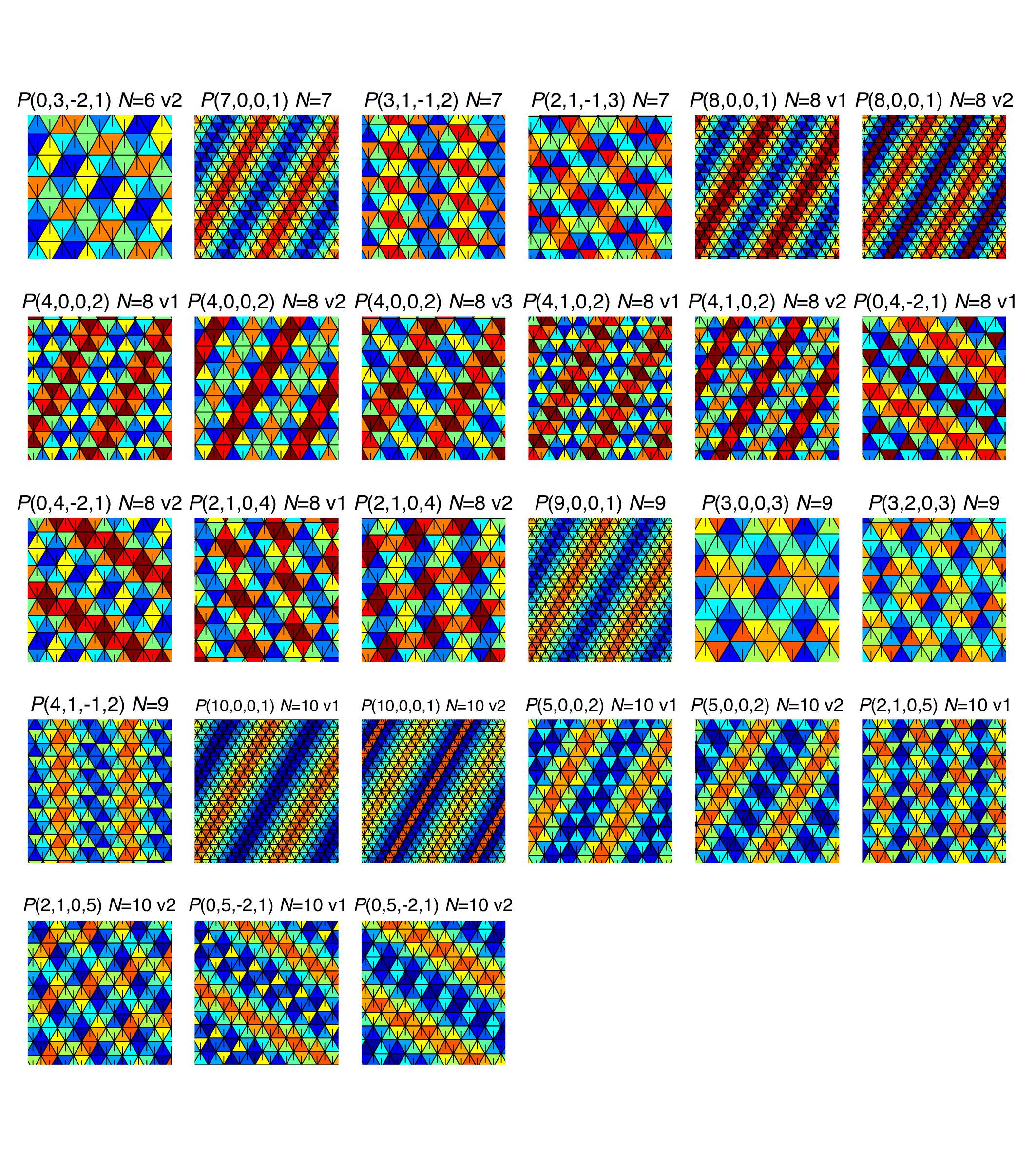}
 \caption{\textbf{List of tilings up to 10 species of triangles (continued).}}
\end{figure}

\begin{figure}[ht]
 \addtocounter{figure}{-1}
 \centering
 \includegraphics[width=0.9\textwidth]{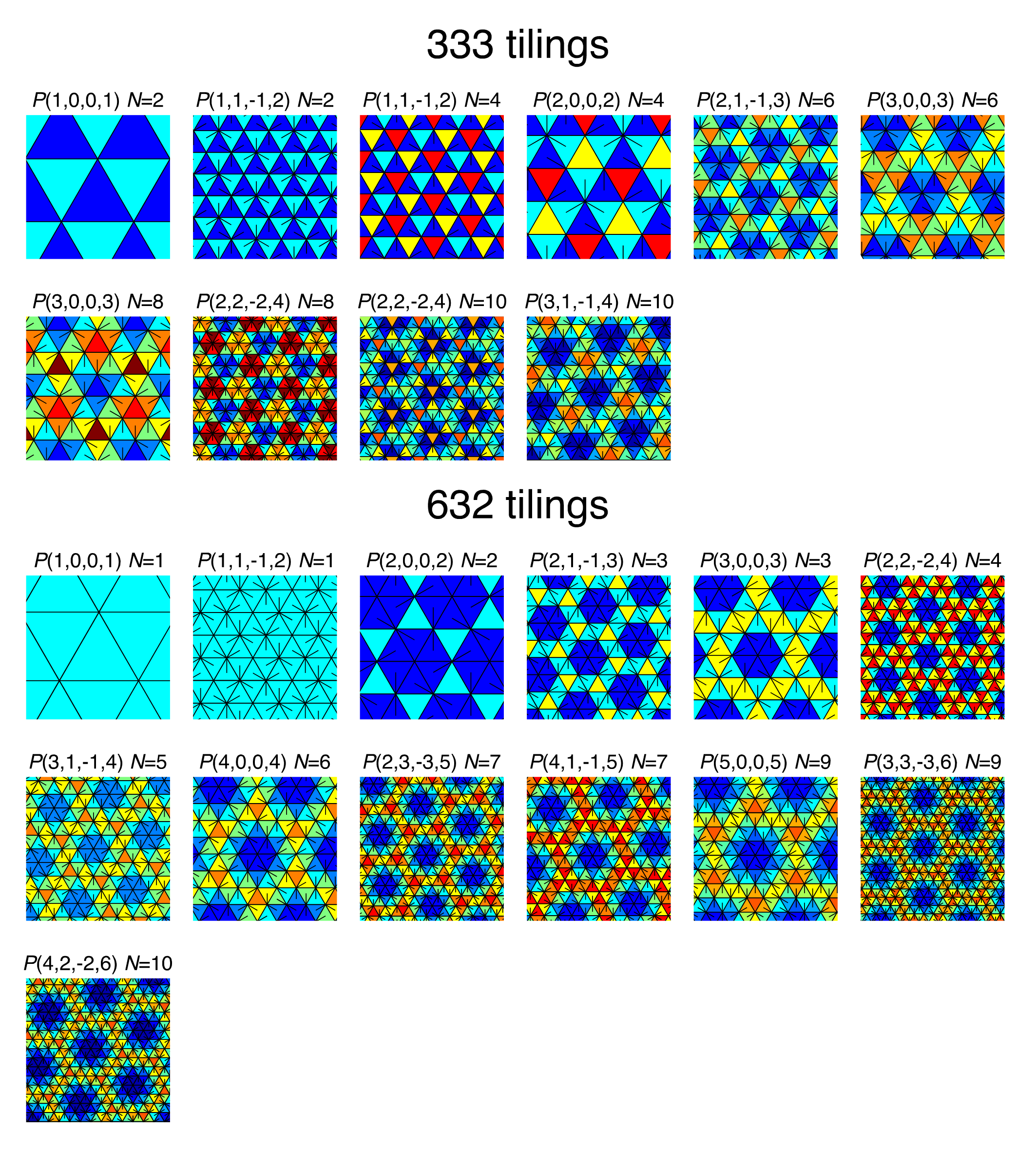}
 \caption{\textbf{List of tilings up to 10 species of triangles (continued).}}
 \label{Sfig:tileList}
\end{figure}


\begin{figure}[ht]
 \centering
 \includegraphics[width=0.99\textwidth]{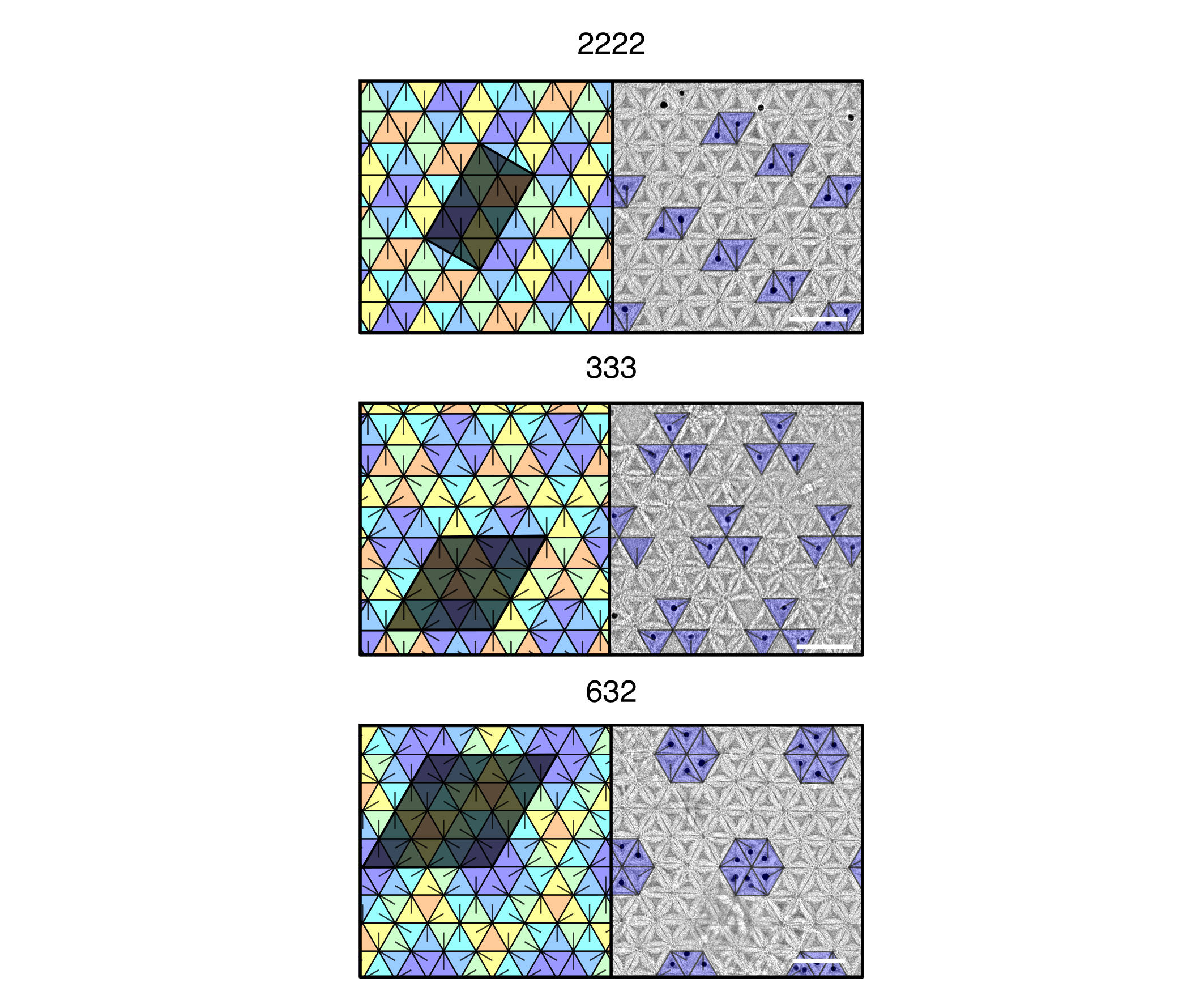}
 \caption{\textbf{Example tilings assembled using 6 species of triangles.}
}
 \label{Sfig:N6}
\end{figure}

\begin{figure}[ht]
 \centering
 \includegraphics[width=0.4\textwidth]{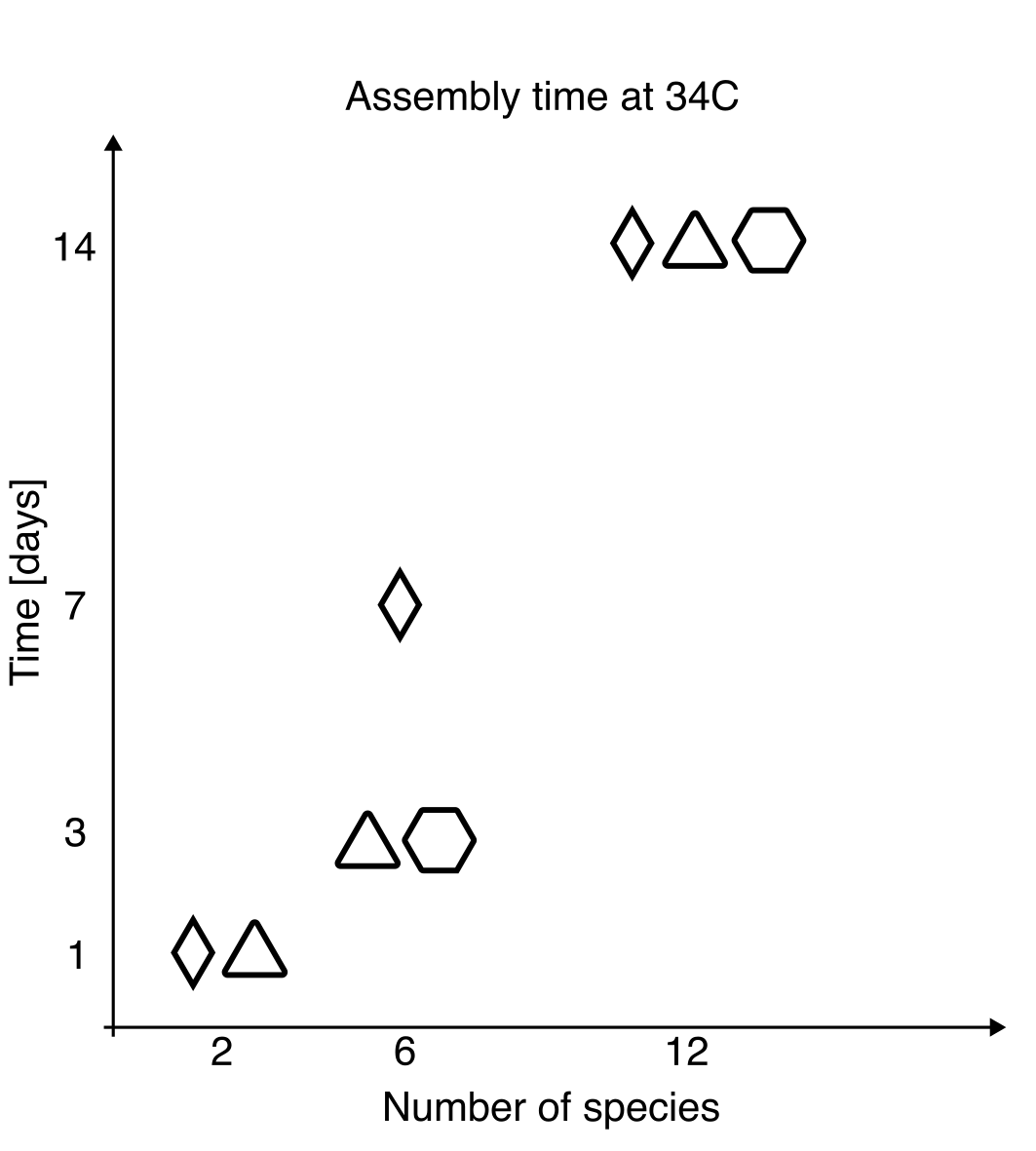}
 \caption{\textbf{Assembly time for various tilings.} Time passed until assembly structures above 500~nm first appear in TEM micrographs, annealed under 34~$^\circ$C. Diamond, triangle, and hexagon denote 2222, 333, and 632 tilings from Fig.~3 and \ref{Sfig:N6}, respectively.
}
 \label{Sfig:tm}
\end{figure}


\begin{figure*}[htb]
 \centering
 \includegraphics[width=1.2\textwidth,angle=90]{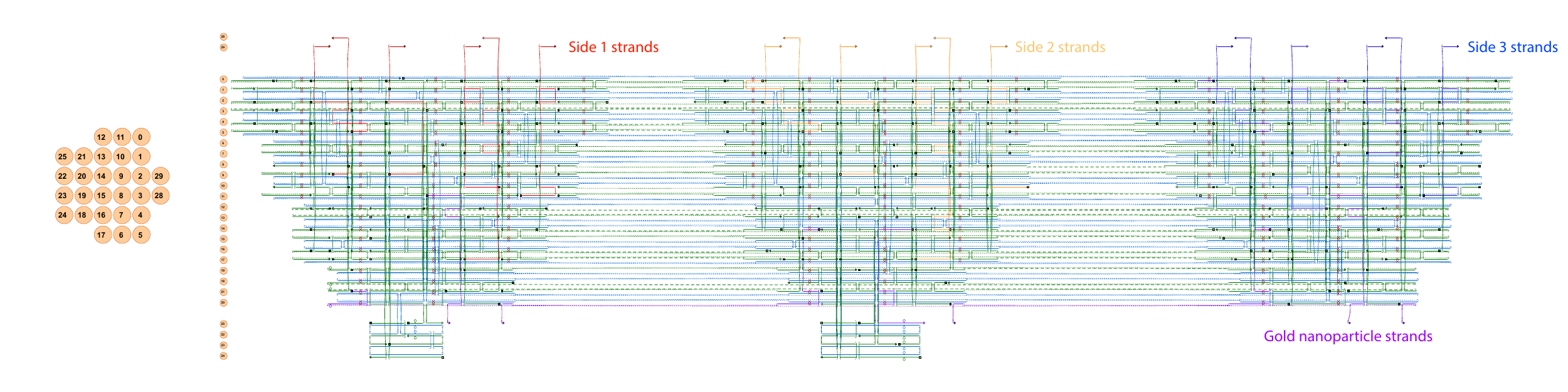}
 \caption{\textbf{Helical numbers and caDNAno designs for the DNA origami triangle.}}
 \label{Sfig:cadnano}
\end{figure*}